\documentclass[preprint,12pt,fleqn]{elsarticle}

\usepackage{amssymb}
\usepackage{amsthm}
\usepackage{amsmath}
\usepackage{graphicx}
\usepackage{hyperref}
\usepackage{multirow}

\journal{Computer Physics Communications}

\begin{document}

\begin{frontmatter}

\title{GPU-Acceleration of the ELPA2 Distributed Eigensolver for Dense Symmetric and Hermitian Eigenproblems}

\author[duke_mems]{Victor Wen-zhe Yu \corref{ca}}
\ead{wenzhe.yu@duke.edu}
\author[molssi]{Jonathan Moussa}
\author[mpcdf,czech]{Pavel K\r{u}s}
\author[mpcdf]{Andreas Marek}
\author[nvidia]{Peter Messmer}
\author[ornl]{Mina Yoon}
\author[mpcdf]{Hermann Lederer}
\author[duke_mems]{Volker Blum}
\cortext[ca]{Corresponding author.}

\address[duke_mems]{Department of Mechanical Engineering and Materials Science, Duke University, Durham, NC 27708, USA}
\address[molssi]{Molecular Sciences Software Institute, Blacksburg, VA 24060, USA}
\address[mpcdf]{Max Planck Computing and Data Facility, Garching D-85747, Germany}
\address[czech]{Institute of Mathematics of the Czech Academy of Sciences, \v{Z}itn\'{a} 25, 115 67 Prague, Czech Republic}
\address[nvidia]{NVIDIA Switzerland, Technoparkstr 1, 8005 Z\"{u}rich, Switzerland}
\address[ornl]{Center for Nanophase Materials Sciences, Oak Ridge National Laboratory, Oak Ridge, TN 37830, USA}

\begin{abstract}
The solution of eigenproblems is often a key computational bottleneck that limits the tractable system size of numerical algorithms, among them electronic structure theory in chemistry and in condensed matter physics. Large eigenproblems can easily exceed the capacity of a single compute node, thus must be solved on distributed-memory parallel computers. We here present GPU-oriented optimizations of the ELPA two-stage tridiagonalization eigensolver (ELPA2). On top of cuBLAS-based GPU offloading, we add a CUDA kernel to speed up the back-transformation of eigenvectors, which can be the computationally most expensive part of the two-stage tridiagonalization algorithm. We benchmark the performance of this GPU-accelerated eigensolver on two hybrid CPU-GPU architectures, namely a compute cluster based on Intel Xeon Gold CPUs and NVIDIA Volta GPUs, and the Summit supercomputer based on IBM POWER9 CPUs and NVIDIA Volta GPUs. Consistent with previous benchmarks on CPU-only architectures, the GPU-accelerated two-stage solver exhibits a parallel performance superior to the one-stage counterpart. Finally, we demonstrate the performance of the GPU-accelerated eigensolver developed in this work for routine semi-local KS-DFT calculations comprising thousands of atoms.
\end{abstract}

\begin{keyword}
Eigensolver \sep dense linear algebra \sep parallel computing \sep high-performance computing \sep GPU \sep CUDA
\end{keyword}

\end{frontmatter}

\section{Introduction}
\label{sec:intro}
Finding the eigenvalues and eigenvectors of large dense matrices is a frequent problem in computational science and engineering. For example, in Kohn-Sham density-functional theory (KS-DFT)~\cite{dft_hohenberg_1964,dft_kohn_1965}, the many-electron problem for the Born-Oppenheimer electronic ground state is reduced to a system of single particle equations that can be discretized into a generalized eigenproblem in the following matrix form
\begin{equation}
\label{eq:gevp}
\boldsymbol{H} \boldsymbol{C} = \boldsymbol{S} \boldsymbol{C} \boldsymbol{\Sigma} .
\end{equation}

\noindent Here the Hamiltonian matrix $\boldsymbol{H}$ and the overlap matrix $\boldsymbol{S}$ are real symmetric or complex Hermitian. $\boldsymbol{S}$ is positive definite. The matrix $\boldsymbol{C}$ and the diagonal matrix $\boldsymbol{\Sigma}$ are the eigenvectors and eigenvalues, respectively, of this eigensystem. In the framework of KS-DFT, $\boldsymbol{C}$ and $\boldsymbol{\Sigma}$ (or the information they carry, at least) are needed for the construction of $\boldsymbol{H}$. Therefore, Eq.~\ref{eq:gevp} is a non-linear problem and must be solved self-consistently. It is possible, and has already been implemented in various codes~\cite{conquest_nakata_2020,fhiaims_blum_2009,onetep_prentice_2020,siesta_garcia_2020}, to restrict the computational cost of the construction of $\boldsymbol{H}$ to scale linearly with respect to the system size $N$ for any semi-local and hybrid exchange-correlation functional. In contrast, the solution of a dense eigenproblem (``diagonalization'') scales as $\mathcal{O}(N^3)$, quickly growing to become prohibitive as $N$ increases to large values.

Today, simulations in materials science and computational chemistry make up a large fraction of supercomputer time used in production (see, e.g., the workload analysis by the National Energy Research Scientific Computing Center~\cite{nersc_2014,nersc_2018}). Particularly, DFT simulations running on high-performance computers are facilitating scientific discoveries across a broad range of disciplines. The progress of this community relies inherently on the availability of scalable solvers on new hardware. In the past, various developments have originated in this community to tackle Eq.~\ref{eq:gevp}~\cite{elsi_yu_2018,elsi_yu_2020}.
\begin{itemize}
\item The algorithm typically employed in a conventional dense eigensolver~\cite{numerical_press_2007,matrix_golub_2013} is tridiagonalization, which brings the original matrix to a tridiagonal form by a series of Householder transformations. This algorithm suffers from the inefficiency of BLAS level-2 matrix-vector operations. New algorithms such as pentadiagonalization~\cite{eigenexa_imamura_2011} and two-stage tridiagonalization~\cite{2stage_lang_1993,2stage_bischof_1994,elpa_auckenthaler_2011,elpa_marek_2014} have been developed, leading to enhanced performance over the conventional one-stage tridiagonalization approach.

\item Iterative eigensolvers~\cite{davidson_davidson_1975,davidson_sleijpen_1996,iterative_payne_1992,iterative_kresse_1996} are commonly employed by DFT codes, particularly those based on plane-wave (PW) basis functions and pseudopotentials. In that case, because of the large number of basis functions, i.e. the large dimension of the matrices in Eq.~\ref{eq:gevp}, needed in an accurate calculation, a direct solution of Eq.~\ref{eq:gevp} is rather infeasible. Iterative eigensolvers are well suited to find a small fraction of low-lying eigenstates, commensurate with the needs of a PW-based code. When using spatially localized basis functions, such as linear combination of atomic orbitals (LCAO), the fraction of needed eigenpairs out of the full matrix dimension can be fairly large. In this scenario, iterative solvers no longer have an advantage over direct eigensolvers.

\item With localized basis functions, locality in the physical system can be translated to sparsity in the $\boldsymbol{H}$ and $\boldsymbol{S}$ matrices. Methods exploiting this sparsity can be formulated as $\mathcal{O}(N) \sim \mathcal{O}(N^2)$~\cite{linear_goedecker_1999,linear_bowler_2012,linear_moussa_2019,feast_polizzi_2009,pexsi_lin_2013,ntpoly_dawson_2018} by circumventing the explicit solution of Eq.~\ref{eq:gevp}. In particular, linear scaling algorithms in a density matrix formalism have been successfully applied to simulations of one million atoms~\cite{million_bowler_2010,million_vandevondele_2012}. Despite the success in extreme-scale simulations, reduced scaling methods come with a computational prefactor that is much larger than that of the $\mathcal{O}(N^3)$ diagonalization method. Moreover, the applicability and optimal performance of reduced scaling methods are often limited to some certain problem types.
\end{itemize}

As of today, dense eigensolvers, with their small computational prefactor and general applicability, remain the default method in most LCAO codes. Even in PW codes, the performance of a dense eigensolver is still crucial, because at some stage of an iterative solver there will typically be a reduced-size dense eigenproblem, the size of which scales with the number of valence electrons in the system being simulated. Therefore, any improvement made to a dense eigensolver would benefit the entire electronic structure community, and the broader field of computational physics in general.

The ubiquitous adoption of graphics processing units (GPUs) in high-performance computing opens up new opportunities to accelerate the solution of dense eigenproblems. A GPU device consists of hundreds to thousands of parallel cores operating at a relatively low frequency. These cores are naturally suited for parallel computational tasks, such as vector and matrix operations found in dense linear algebra. On top of that, GPUs typically have a power efficiency superior to traditional central processing units (CPUs), and therefore play an important role in supercomputing towards the exascale. According to the November 2020 release of the TOP500 list~\cite{top500}, six of the top ten machines have GPU accelerators, including Summit, the world's second fastest computer (was the fastest from June 2018 to June 2020) based on IBM POWER9 CPUs and NVIDIA Tesla Volta V100 GPUs.

GPU-accelerated eigensolvers have long been available in GPU-oriented linear algebra packages such as cuSOLVER (single GPU)~\cite{cusolver}, cuSOLVER-MG (multiple GPUs)~\cite{cusolver}, and MAGMA (CPUs and multiple GPUs)~\cite{magma_tomov_2010,magma_dongarra_2014}. These packages are designed and optimized for shared-memory host architectures. They can be very fast, but the problem size they can tackle is limited by the memory capacity of a single compute node. Fully exploiting the power of GPU-accelerated supercomputers would require a distributed-memory implementation.

The MPI-parallel, distributed-memory ELPA library~\cite{elpa} implements the conventional one-stage diagonalization method and the two-stage diagonalization proposed in Refs.~\cite{2stage_lang_1993,2stage_bischof_1994}, known as the ``ELPA1'' and ``ELPA2'' solvers, respectively. The ELPA library is being used in a large number of quantum chemistry and solid state physics software packages (mentioned either in the publications cited here or in the documentation of the package), including ABINIT~\cite{abinit_gonze_2020}, BerkeleyGW~\cite{berkeleygw_deslippe_2012}, CP2K~\cite{cp2k_kuhne_2020}, CPMD~\cite{cpmd_kloffel_2021}, DFTB+~\cite{dftb_hourahine_2020}, FHI-aims~\cite{fhiaims_blum_2009}, GPAW~\cite{gpaw_enkovaara_2010}, NWChem~\cite{nwchem_apra_2020}, Octopus~\cite{octopus_tancognedejean_2020}, OpenMX~\cite{openmx_ozaki_2003}, QuantumATK~\cite{quantumatk_smidstrup_2019}, Quantum ESPRESSO~\cite{qe_giannozzi_2020}, SIESTA~\cite{siesta_garcia_2020}, VASP~\cite{vasp_kresse_1996}, and WIEN2k~\cite{wien2k_blaha_2020}. Both the ELPA1 and ELPA2 solvers have been ported to GPUs by substituting BLAS calls with the corresponding cuBLAS functions~\cite{elpa_kus_2019a}, making ELPA the first publicly available, distributed-memory, GPU-accelerated eigensolver to our knowledge. We note the ongoing effort of the SLATE project~\cite{slate}, a more general distributed-memory dense linear algebra framework being developed to replace ScaLAPACK on modern high-performance computing (HPC) machines with large numbers of CPU cores and accelerators. The eigensolver implementation in SLATE adopts the same two-stage diagonalization method as implemented in ELPA2, but does not return eigenvectors at the time of writing (September 2020).

For brevity, the CPU-only version of ELPA1 and ELPA2 and the GPU-accelerated version of ELPA1 and ELPA2 are hereafter referred to as CPU-ELPA1, CPU-ELPA2, GPU-ELPA1, and GPU-ELPA2, respectively. This paper presents the first complete overview of GPU-ELPA2, including in-depth benchmarks. The only previously published benchmark data of GPU-ELPA2 were restricted to single-node tests of a preliminary version of GPU-ELPA2, shown as a single line of data in Fig.~5 of Ref.~\cite{elpa_kus_2019b}. When using two IBM POWER8 CPUs (24 cores in total) and four NVIDIA Pascal P100 GPUs, GPU-ELPA1 delivers up to 11.9x performance boost compared to CPU-ELPA1 using 24 CPU cores. In this regime, the performance of GPU-ELPA1 was still better than that of GPU-ELPA2. The extensive set of benchmarks in the present paper demonstrate the regimes in which the present version of GPU-ELPA2 provides a significant advantage. Historically, the GPU port of ELPA2 has its roots in 2013, when Peter Messmer of NVIDIA programmed the first proof-of-concept version of GPU-ELPA2. Then the code was refactored and merged into the mainline version of ELPA, and has been available in released versions of the ELPA eigensolver library since 2016. In this paper, we report the current complete version of GPU-ELPA2, including our latest optimizations and developments that enable a performance improvement on distributed-memory, GPU-accelerated architectures. Specifically, several synchronizations and memory transfers between CPUs and GPUs have been optimized. Additionally, some kernels in one of the major computational steps, the tridiagonal-to-banded back-transformation of eigenvectors (Eq.~\ref{eq:2stage_bkwd1}), have been rewritten.

The rest of this paper is organized as follows. First, we briefly review the two-stage diagonalization algorithm, in particular the tridiagonal-to-banded back-transformation of eigenvectors and its CPU implementation in the ELPA library. Next, we outline the GPU acceleration strategies employed in GPU-ELPA2, and elaborate on our CUDA implementation of the tridiagonal-to-banded back-transformation of eigenvectors, which is essentially a GPU extension of the algorithm in Refs.~\cite{elpa_auckenthaler_2011,elpa_auckenthaler_2013}. We then benchmark the performance and scalability of the GPU-ELPA2 solver on two GPU-accelerated computers, namely the Talos cluster at Max Planck Computing and Data Facility in Garching, Germany, based on Intel Xeon Gold CPUs and NVIDIA Volta GPUs, and the Summit supercomputer at Oak Ridge National Laboratory in Tennessee, USA, based on IBM POWER9 CPUs and NVIDIA Volta GPUs. Finally, we demonstrate the performance of the GPU-ELPA2 solver for practical computational physics problems on the current top supercomputer, Summit, for routine semi-local KS-DFT calculations including thousands of atoms without sacrificing any accuracy.

\section{Two-Stage Tridiagonalization in ELPA2}
\label{sec:review_2stage}
\subsection{Overview of the Two-Stage Tridiagonalization}
\label{subsec:cpu_2stage}
The textbook procedure~\cite{numerical_press_2007,matrix_golub_2013} to solve a dense generalized eigenproblem, like the one in Eq.~\ref{eq:gevp}, first computes the Cholesky factorization of $\boldsymbol{S}$
\begin{equation}
\label{eq:cholesky}
\boldsymbol{S} = \boldsymbol{L} \boldsymbol{L}^* ,
\end{equation}

\noindent then uses $\boldsymbol{L}$ to transform Eq.~\ref{eq:gevp} to a standard eigenproblem
\begin{equation}
\label{eq:evp}
\boldsymbol{\tilde{H}} \boldsymbol{\tilde{C}} = \boldsymbol{\tilde{C}} \boldsymbol{\Sigma} .
\end{equation}

\noindent $\boldsymbol{\tilde{H}}$ is
\begin{equation}
\label{eq:gevp2evp}
\boldsymbol{\tilde{H}} = \boldsymbol{L}^{-1} \boldsymbol{H} (\boldsymbol{L}^*)^{-1} ,
\end{equation}

\noindent and the eigenvectors $\boldsymbol{\tilde{C}}$ must be back-transformed in order to retrieve the eigenvectors of Eq.~\ref{eq:gevp}, i.e.
\begin{equation}
\label{eq:evp2gevp}
\boldsymbol{C} = (\boldsymbol{L}^*)^{-1} \boldsymbol{\tilde{C}} .
\end{equation}

The direct solution of Eq.~\ref{eq:evp} is based on tridiagonalization, that is, the full matrix $\boldsymbol{\tilde{H}}$ is transformed to a tridiagonal matrix $\boldsymbol{T}$. This is typically accomplished by individual Householder transformations, which take the shape of matrix-vector operations. Eigenvalues and eigenvectors of $\boldsymbol{T}$ can be easily (compared to the original problem) solved. Then, eigenvectors of $\boldsymbol{T}$ are back-transformed to obtain eigenvectors of $\boldsymbol{\tilde{H}}$. This algorithm is adopted by a variety of dense linear algebra packages, such as LAPACK~\cite{lapack_anderson_1999} targeting sequential and shared-memory parallel architectures; cuSOLVER~\cite{cusolver} and MAGMA~\cite{magma_tomov_2010,magma_dongarra_2014} targeting shared-memory architectures with GPU accelerators; ScaLAPACK~\cite{scalapack_blackford_1997} and Elemental~\cite{elemental_poulson_2013} targeting distributed-memory parallel architectures. However, as mentioned above, the tridiagonalization step makes extensive use of memory-bound, BLAS level-2 matrix-vector operations, whose performance is limited on modern computer architectures.

\begin{figure*}[ht!]
\centering
\includegraphics[width=\textwidth]{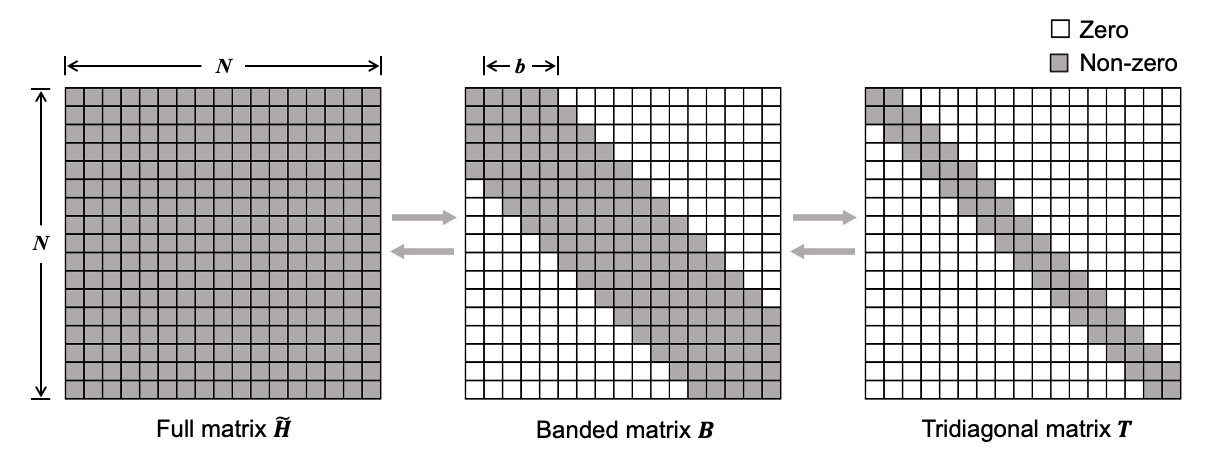}
\caption{Computational steps of the two-stage tridiagonalization approach: the reduction of the full matrix $\boldsymbol{\tilde{H}}$ to a banded matrix $\boldsymbol{B}$, the reduction of the banded matrix $\boldsymbol{B}$ to a tridiagonal matrix $\boldsymbol{T}$, the solution of the tridiagonal eigenproblem, the back-transformation of the eigenvectors to the banded form, and the back-transformation of the eigenvectors to the full form. Matrix size $N=17$. Semi-bandwidth $b=4$.}
\label{fig:2stage}
\end{figure*}

The two-stage tridiagonalization algorithm proposed by Bischof, Sun, and Lang~\cite{2stage_lang_1993,2stage_bischof_1994} is an established alternative to the conventional one-stage method. As shown in Eq.~\ref{eq:2stage} below and further illustrated in Fig.~\ref{fig:2stage}, the tridiagonalization of the full matrix $\boldsymbol{\tilde{H}}$ is carried out in two transformations. The first transformation $\boldsymbol{P}$ reduces $\boldsymbol{\tilde{H}}$ to a banded matrix $\boldsymbol{B}$, and the second transformation $\boldsymbol{Q}$ reduces $\boldsymbol{B}$ to a tridiagonal matrix $\boldsymbol{T}$. The eigenvalues $\boldsymbol{\Sigma}$ and eigenvectors $\boldsymbol{X}$ of $\boldsymbol{T}$ are solved as done in the one-stage method. The back-transformation of eigenvectors is also carried out in two steps. $\boldsymbol{X}$ is first back-transformed to $\boldsymbol{Y}$, the eigenvectors of $\boldsymbol{B}$, then to $\boldsymbol{\tilde{C}}$, the eigenvectors of $\boldsymbol{\tilde{H}}$.
\begin{subequations}
\label{eq:2stage}
\begin{align}
\boldsymbol{B} & = \boldsymbol{P} \boldsymbol{\tilde{H}} \boldsymbol{P}^* , \label{eq:2stage_fwd1} \\
\boldsymbol{T} & = \boldsymbol{Q} \boldsymbol{B} \boldsymbol{Q}^* , \label{eq:2stage_fwd2} \\
\boldsymbol{T} \boldsymbol{X} & = \boldsymbol{X} \boldsymbol{\Sigma} , \label{eq:2stage_solve} \\
\boldsymbol{Y} & = \boldsymbol{Q}^* \boldsymbol{X} , \label{eq:2stage_bkwd1} \\
\boldsymbol{\tilde{C}} & = \boldsymbol{P}^* \boldsymbol{Y} . \label{eq:2stage_bkwd2}
\end{align}
\end{subequations}

This two-stage tridiagonalization approach is implemented in several linear algebra software packages~\cite{elpa_auckenthaler_2011,elpa_marek_2014,magma_tomov_2010,slate,lapack_anderson_1999,mkl_arturov_2018}, including a high-performance distributed-memory implementation in the ELPA library~\cite{elpa_auckenthaler_2011,elpa_marek_2014}. The introduction of the banded matrix stage leads to faster computation compared to the one-stage method, for the transformation in Eq.~\ref{eq:2stage_fwd1} mostly involves highly efficient BLAS level-3 matrix-matrix operations, and the transformation in Eq.~\ref{eq:2stage_fwd2} only works on a sparse banded matrix $\boldsymbol{B}$ instead of a full matrix. The solution of Eq.~\ref{eq:2stage_solve} is accelerated in ELPA by extending the divide-and-conquer symmetric tridiagonal eigensolver~\cite{dc_cuppen_1980,dc_gu_1995,dc_tisseur_1999} such that unwanted eigenvectors are not computed~\cite{elpa_auckenthaler_2011,elpa_auckenthaler_2013}. Regarding the back-transformation of eigenvectors, due to the data layout, it is rather difficult to efficiently use BLAS level-3 routines for Eq.~\ref{eq:2stage_bkwd1}~\cite{elpa_auckenthaler_2011,elpa_kus_2019b,elpa_auckenthaler_2013}, which involves the application of a sequence of short Householder transformations as detailed in Sec.~\ref{subsubsec:cpu_step4}. Manually optimized ``kernels'' written in architecture-specific instruction sets are employed for this particular step. The ELPA2 solver is highly scalable on massively parallel, distributed-memory architectures. It avoids global MPI communications by using a 2-dimensional (2D) process grid and restricting the communication to take place within either the row direction or the column direction. Input, output, and most of the internal matrices are distributed following the standard 2D block-cyclic distribution scheme. Depending on the size of the problem, ELPA2 scales to at least tens of thousands of CPU cores~\cite{elpa_marek_2014,elpa_gutheil_2014,elpa_cook_2018}.

\subsection{Tridiagonal-to-Banded Back-Transformation of Eigenvectors}
\label{subsec:cpu_step4}
We now summarize the algorithms of the banded-to-tridiagonal transformation (Eq.~\ref{eq:2stage_fwd2}) and the tridiagonal-to-banded back-transformation (Eq.~\ref{eq:2stage_bkwd1}), and their CPU implementation in ELPA2. The reader is also referred to Refs.~\cite{2stage_lang_1993,2stage_bischof_1994,elpa_auckenthaler_2011,elpa_auckenthaler_2013}. Throughout this and the following sections, we will use the matrix in Fig.~\ref{fig:2stage}, with matrix size $N=17$ and semi-bandwidth $b=4$, to illustrate the various algorithms. It is important to note that in a typical GPU-accelerated calculation, $N$ is at least several thousand, and $b$ is a multiple of 32 in our implementation. The small example in Fig.~\ref{fig:2stage} is chosen only because a larger one would be difficult to visualize.

\subsubsection{Banded-to-Tridiagonal Transformation}
\label{subsubsec:cpu_step2}
ELPA2 relies on the ``bulge chasing'' algorithm~\cite{2stage_lang_1993,2stage_bischof_1994} to reduce a banded matrix $\boldsymbol{B}$ to a tridiagonal matrix $\boldsymbol{T}$. Let $N$ and $b$ denote the dimension and semi-bandwidth of $\boldsymbol{B}$, respectively. The banded-to-tridiagonal transformation is done in $N-2$ stages, with $(N-i)/b$ sweeps in the $i^\text{th}$ stage. The first sweep of the $i^\text{th}$ stage reduces the $i^\text{th}$ column of $\boldsymbol{B}$ to the target tridiagonal form, at the same time introducing fill-ins (``bulges'') to the remainder of $\boldsymbol{B}$. From the second sweep on, the first column of the fill-ins introduced in the previous sweep is eliminated, while introducing new fill-ins further down the matrix.

\begin{figure*}[ht!]
\centering
\includegraphics[width=\textwidth]{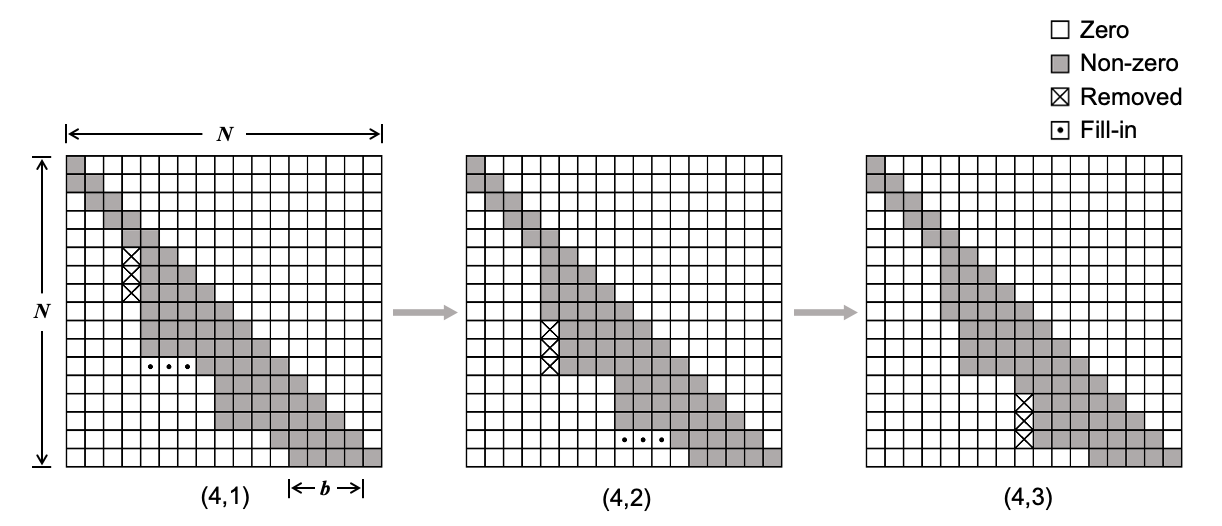}
\caption{Visualization of the sweeps in the fourth stage of the bulge chasing procedure for matrix size $N=17$ and semi-bandwidth $b=4$. The $(i,j)$ pair denotes the stage and sweep indexes. For symmetric and Hermitian eigenproblems, only the lower triangular part is considered.}
\label{fig:step2}
\end{figure*}

This procedure is visualized in Fig.~\ref{fig:step2} for $N=17$ and $b=4$, with the fourth stage as an example. The first three columns of the matrix have been transformed to the target tridiagonal form by the previous three stages. The fourth stage consists of $(N-i)/b=3$ sweeps. As shown in Fig.~\ref{fig:step2}, in the first sweep of the fourth stage, the fourth column is brought to the desired tridiagonal form by a Householder transformation that eliminates three elements from the fourth column and at the same time introduces three fill-ins. The second and third sweeps partly eliminate the fill-ins introduced in the previous sweeps. The remaining fill-ins are eliminated in the next stages.

\begin{figure*}[ht!]
\centering
\includegraphics[width=0.6\textwidth]{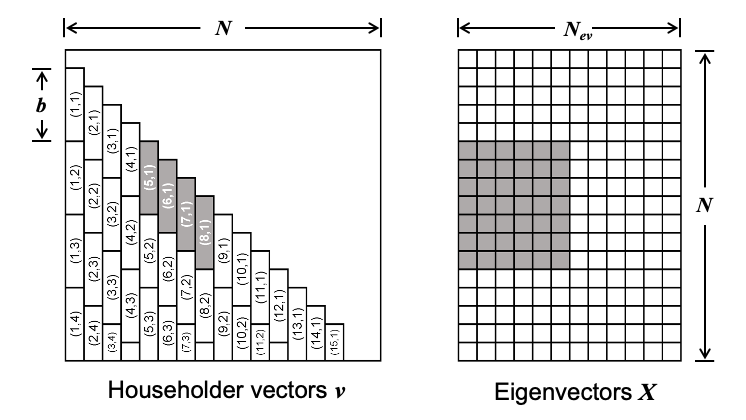}
\caption{Visualization of Householder vectors $\boldsymbol{v}_{(i,j)}$ and eigenvectors $\boldsymbol{X}$ involved in the tridiagonal-to-banded back-transformation. Matrix size $N=17$. Semi-bandwidth $b=4$. Number of eigenvectors $N_{ev}=12$. The left panel shows where vectors $\boldsymbol{v}_{(i,j)}$ are generated in the banded-to-tridiagonal transformation. Note that in the code these vectors are stored in a separate array. They cannot be stored in place, because the transformations generated in the full-to-banded transformation are stored there. The $(i,j)$ pair inside a vector denotes its stage and sweep indexes. For symmetric and Hermitian eigenproblems, only the lower triangular part is considered. The right panel shows the eigenvectors $\boldsymbol{X}$ to be back-transformed. When using multiple MPI processes, the local computation performed by each process is the application of a series of Householder transformations $\boldsymbol{Q}_{(i,j)}$ to a part of the eigenvectors $\boldsymbol{X}$. As a fictitious example, in the first step of the back-transformation, an MPI process could be responsible for applying the shaded Householder transformations sequentially, i.e. (8,1) $\rightarrow$ (7,1) $\rightarrow$ (6,1) $\rightarrow$ (5,1), to the shaded part of the eigenvectors.}
\label{fig:step4}
\end{figure*}

The Householder transformation applied in the $j^\text{th}$ sweep of the $i^\text{th}$ stage can be written as
\begin{equation}
\label{eq:householder}
\boldsymbol{Q}_{(i,j)} = \boldsymbol{Q}^*_{(i,j)} = \boldsymbol{I} - \tau_{(i,j)} \boldsymbol{v}_{(i,j)} \boldsymbol{v}_{(i,j)}^* ,
\end{equation}

\noindent where $\boldsymbol{Q}_{(i,j)}$ is a Householder transformation matrix; $\boldsymbol{I}$ is the identity matrix; the scalar $\tau_{(i,j)}$ and vector $\boldsymbol{v}_{(i,j)}$ are computed following the standard Householder method~\cite{numerical_press_2007,householder_householder_1958}. Most Householder vectors $\boldsymbol{v}_{(i,j)}$ have a length equal to $b$, except that $\boldsymbol{v}_{(i,j)}$ generated in the last sweep of each stage may be shorter. Matrix-vector operations are still needed in order to apply these Householder transformations, but the computational cost is much less than in the one-stage tridiagonalization algorithm, as the vectors are much shorter. The left panel of Fig.~\ref{fig:step4} shows the shape of the Householder vectors $\boldsymbol{v}_{(i,j)}$. In the actual code, the Householder matrices $\boldsymbol{Q}_{(i,j)}$ are never explicitly constructed. Instead, $\tau_{(i,j)}$ and $\boldsymbol{v}_{(i,j)}$ are stored and used for the transformations in Eqs.~\ref{eq:2stage_fwd2} and \ref{eq:2stage_bkwd1}, where $\boldsymbol{Q}$ is the product of all $\boldsymbol{Q}_{(i,j)}$.

\subsubsection{Tridiagonal-to-Banded Back-Transformation of Eigenvectors}
\label{subsubsec:cpu_step4}
After the tridiagonal eigenproblem (Eq.~\ref{eq:2stage_solve}) is solved, we get the eigenvalues $\boldsymbol{\Sigma}$ and eigenvectors $\boldsymbol{X}$ of the tridiagonal matrix $\boldsymbol{T}$. The main task of the tridiagonal-to-banded back-transformation is to apply all $\boldsymbol{Q}^*_{(i,j)}$ to $\boldsymbol{X}$ (Eq.~\ref{eq:2stage_bkwd1}). It is obvious that the eigenvectors can be back-transformed independently, leading to a trivial parallelism over eigenvectors. The Householder transformations $\boldsymbol{Q}^*_{(i,j)}$ cannot be applied simultaneously, for several transformations may update the same part of an eigenvector. The order in which the Householder transformations are applied in the tridiagonal-to-banded back-transformation must be the reverse of the order in which they are generated in the banded-to-tridiagonal transformation. However, as the Householder vectors $\boldsymbol{v}_{(i,j)}$ are shorter than the eigenvectors $\boldsymbol{X}$, a single transformation $\boldsymbol{Q}^*_{(i,j)}$ only alters a few rows of $\boldsymbol{X}$ where $\boldsymbol{v}_{(i,j)}$ has non-zero values. For instance, the Householder vector $\boldsymbol{v}_{(1,4)}$ at the left bottom corner of Fig.~\ref{fig:step4} only alters the last four rows of $\boldsymbol{X}$. This leads to another level of parallelism to be exploited within each individual eigenvector.

In ELPA2, the $N_p$ MPI processes are organized in an $N_{pr}$ by $N_{pc}$ grid. The $N_{ev}$ eigenvectors are uniformly distributed across the $N_{pc}$ process columns. Within a process column, the eigenvectors are distributed across the $N_{pr}$ processes in a block manner. Each MPI process applies a series of Householder transformations to its local part of the eigenvector matrix $\boldsymbol{X}$. The shaded part in Fig.~\ref{fig:step4} shows an example of the local computation by a process. This process applies four Householder transformations, $\boldsymbol{Q}^*_{(8,1)}$, $\boldsymbol{Q}^*_{(7,1)}$, $\boldsymbol{Q}^*_{(6,1)}$, and $\boldsymbol{Q}^*_{(5,1)}$, to its local part of the eigenvector matrix, referred to as $\boldsymbol{X}_\text{local}$ hereafter. Apparently, $\boldsymbol{Q}^*_{(4,1)}$, $\boldsymbol{Q}^*_{(3,1)}$, and $\boldsymbol{Q}^*_{(2,1)}$ also modify the top part of $\boldsymbol{X}_\text{local}$. These three transformations are however applied by the upper adjacent MPI process in the same process column. Therefore, the top three rows of $\boldsymbol{X}_\text{local}$ must be exchanged with the upper adjacent process. Likewise, the bottom three rows of $\boldsymbol{X}_\text{local}$ must be exchanged with the lower adjacent process in the same process column. The middle part of $\boldsymbol{X}_\text{local}$ is not involved in any data exchange. It appears in Fig.~\ref{fig:step4} that almost the entire $\boldsymbol{X}_\text{local}$ matrix needs to be exchanged between adjacent process rows, as there is only one row in the middle part. However, the height of the middle part actually increases with the matrix size $N$, whereas the height of the top and bottom parts can never exceed the semi-bandwidth $b$. Therefore, the amount of data that needs to be exchanged, i.e., data in the top and bottom parts, is limited by $b$, usually accounting for a small fraction of $\boldsymbol{X}_\text{local}$.

\section{GPU Acceleration of ELPA2}
\label{sec:gpu_2stage}
The two-stage tridiagonalization is implemented as five separate subroutines in ELPA2, corresponding to the five steps in Eq.~\ref{eq:2stage}. The input, output, and internal working matrices of ELPA2 are distributed across CPU cores. ELPA2 relies on its own explicit MPI calls to handle distributed linear algebra operations. The efficient MPI communication pattern in the CPU version of ELPA2 is not altered in the GPU version, where GPU acceleration is mainly realized by substituting local BLAS calls with the corresponding cuBLAS functions, as is done for the full-to-banded transformation (Eq.~\ref{eq:2stage_fwd1}), the solution of the tridiagonal eigenproblem (Eq.~\ref{eq:2stage_solve}), the banded-to-full back-transformation (Eq.~\ref{eq:2stage_bkwd2}), and additionally the matrix multiplications in the transformation between generalized and standard eigenproblems (Eqs.~\ref{eq:gevp2evp} and \ref{eq:evp2gevp}). The tridiagonal-to-banded back-transformation (Eq.~\ref{eq:2stage_bkwd1}) is GPU accelerated by a CUDA implementation of Eq.~\ref{eq:householder}. The banded-to-tridiagonal transformation (Eq.~\ref{eq:2stage_fwd2}) has not been ported to GPUs, because of its low computational cost as shown in Fig.~\ref{fig:decomp5_summit}.

\subsection{GPU Offloading via cuBLAS and Multi-Process Service}
\label{subsec:gpu_cublas}
The API of cuBLAS is designed to be almost identical to that of the standard CPU BLAS, making cuBLAS-based GPU offloading straightforward. Here we only comment on two technical aspects.

First, the communication and synchronization between CPUs and GPUs should be avoided as much as possible. Before calling a cuBLAS function, the input arrays must reside on the GPU memory, which often requires a copy of the data from CPU to GPU. In order for the CPU to access the result of cuBLAS, another copy from GPU to CPU is needed. In our GPU porting of ELPA2, CPU-GPU memory transfers are reduced to minimum by leaving data on the GPU as long as possible. Most often, GPU data is copied back to the CPU to participate in an MPI communication. In the version discussed in this paper, we have not yet explored GPU-aware MPI to directly communicate data on the GPU. In each of the GPU-accelerated subroutines, the allocation of GPU memory, which implies a CPU-GPU synchronization, is performed before the main work begins by precomputing the size of the allocation. The allocated GPU memory is reused wherever possible, and is deallocated after the main work finishes. Avoiding frequent allocation and deallocation of GPU memory in a loop or in a recursive routine helps reduce the amount of implicit CPU-GPU synchronization.

Second, the CPU code of the GPU-accelerated version of ELPA2 operates in a pure MPI mode without threading, i.e., one MPI process for each CPU core. As most (if not all) mainstream computers today have more CPU cores than GPU devices, several MPI processes would have to share one GPU device. When the size of an eigenproblem is relatively small, the amount of work assigned to each individual MPI process may not be able to saturate the GPU. In such cases, the NVIDIA Multi-Process Service (MPS) transparently allows work from different MPI processes to be executed concurrently on the GPU~\cite{mps}. Specifically, with MPS, memory transfer and computation requests from different MPI processes are automatically overlapped whenever possible, increasing the overall GPU utilization. Without MPS, memory transfer and computation requests from different MPI processes are not overlapped on a single GPU. Multiple MPI processes will still be able to use the same GPU even if MPS is not switched on, just not concurrently, degrading performance slightly (see Fig.~\ref{fig:all_talos} in Sec.~\ref{subsec:results_overall} for an example). It is thus recommended to use ELPA with MPS switched on.

We note that MPS does not apply only to cuBLAS but also to other, kernel-based GPU offloading in our implementation.

\subsection{CUDA Implementation of Parallel Householder Transformations}
\label{subsec:gpu_step4}
In this section, we present our CUDA implementation of the local computation in Fig.~\ref{fig:step4}, which is the key step in the tridiagonal-to-banded back-transformation (Eq.~\ref{eq:2stage_bkwd1}). An MPI process is responsible for applying $N_v$ Householder transformations to its local eigenvector matrix $\boldsymbol{X}_\text{local}$, which has $N_R$ rows and $N_C$ columns. A Householder transformation defined by $\tau$ and $\boldsymbol{v}$ is applied to an eigenvector $\boldsymbol{x}$ by
\begin{equation}
\label{eq:kernel}
(\boldsymbol{I} - \tau \boldsymbol{v} \boldsymbol{v}^*) \boldsymbol{x} = \boldsymbol{x} - \tau \boldsymbol{v} (\boldsymbol{v}^* \boldsymbol{x}) .
\end{equation}

\noindent As explained in Sec.~\ref{subsec:cpu_step4}, the $N_v$ Householder transformations must be applied in a strict order, which is the reverse of the order in which they are generated in the banded-to-tridiagonal transformation. Consider the example in Fig.~\ref{fig:step4} again, the first transformation would be $\boldsymbol{Q}^*_{(8,1)}$, then $\boldsymbol{Q}^*_{(7,1)}$, $\boldsymbol{Q}^*_{(6,1)}$, and $\boldsymbol{Q}^*_{(5,1)}$. It is obvious that from one transformation to the next, the rows of $\boldsymbol{X}_\text{local}$ modified by the transformation are shifted upward by one row. This is also seen in Fig.~\ref{fig:step4}, where $\boldsymbol{Q}^*_{(8,1)}$ modifies the fourth to seventh rows of $\boldsymbol{X}_\text{local}$, $\boldsymbol{Q}^*_{(7,1)}$ modifies the third to sixth rows of $\boldsymbol{X}_\text{local}$, and so forth. Each individual $\boldsymbol{Q}^*_{(i,j)}$ actually only modifies $b$ rows of $\boldsymbol{X}_\text{local}$.

In a CUDA program, the large number of GPU cores are arranged into a grid of blocks, each of which in turn comprises a grid of threads. All the GPU cores work in a single instruction, multiple thread (SIMT) fashion, i.e., a single instruction is simultaneously executed by multiple threads with different data~\cite{cuda_nickolls_2008}. In order to map the local computation in Fig.~\ref{fig:step4} to the GPU cores, we choose a 1D block grid with a 1D thread grid within each block. The CUDA kernel is launched with $N_C$ blocks and $b$ threads per block. $N_C$, the number of eigenvectors to be back-transformed on a given MPI process, is proportional to the size of the eigenproblem. For example, if there are 40,000 eigenvectors (smallest example in Fig.~\ref{fig:all_summit}) and one node on Summit (42 MPI processes in a layout of $7 \times 6$ process rows and columns), $N_C$ is $6,667$. Each block of threads works on an eigenvector, and each thread works on an element of this eigenvector. Specifically, the Householder transformations are applied to $\boldsymbol{X}_\text{local}$ as follows:
\begin{enumerate}
\item Copy $\boldsymbol{X}_\text{local}$, as well as all $\boldsymbol{v}$ and $\tau$ that ever update $\boldsymbol{X}_\text{local}$, from CPU to GPU. Each block of GPU threads is responsible for one column of $\boldsymbol{X}_\text{local}$, denoted as $\boldsymbol{x}$. The $N_C$ columns of $\boldsymbol{X}_\text{local}$ are processed by the $N_C$ blocks simultaneously. Steps 2--5 below correspond to the workflow of one block.

\item Compute Eq.~\ref{eq:kernel} for the first Householder transformation.
\begin{enumerate}
\item Compute dot product $\boldsymbol{v}^* \boldsymbol{x}$. Only $b$ elements of $\boldsymbol{x}$ contribute to the dot product. Each of the $b$ GPU threads loads one element of $\boldsymbol{v}$ and one element of $\boldsymbol{x}$, multiplies them together, and stores the thread-local result in shared memory. The final dot product is obtained by a parallel reduction involving all the threads in a block.

\item Update $\boldsymbol{x}$ by $\boldsymbol{x} = \boldsymbol{x} - \tau (\boldsymbol{v}^* \boldsymbol{x}) \boldsymbol{v}$. Again, only $b$ elements of $\boldsymbol{x}$ are modified. Since the dot product $\boldsymbol{v}^* \boldsymbol{x}$ has already been computed and $\tau$ is a scalar, this update can be done in a straightforward element-wise fashion, i.e., each thread updates one element of $\boldsymbol{x}$.
\end{enumerate}

\item Among the $b$ elements of $\boldsymbol{x}$ that are updated by the first Householder transformation, only the lowest element will not be affected by the next Householder transformation. Before applying the next transformation, the last thread in each block writes its element of $\boldsymbol{x}$ into $\boldsymbol{X}_\text{local}$. Then thread $i$ ($i \ge 1$) takes the element of $\boldsymbol{x}$ from thread ($i-1$), while thread 0 loads a new element from $\boldsymbol{X}_\text{local}$.

\item Now all threads are ready to repeat steps 1, 2, and 3 for the next Householder transformation.

\item The CUDA kernel finishes when all transformations are applied. The updated $\boldsymbol{X}_\text{local}$ is then copied back to CPU.
\end{enumerate}

\begin{figure*}[ht!]
\centering
\includegraphics[width=0.85\textwidth]{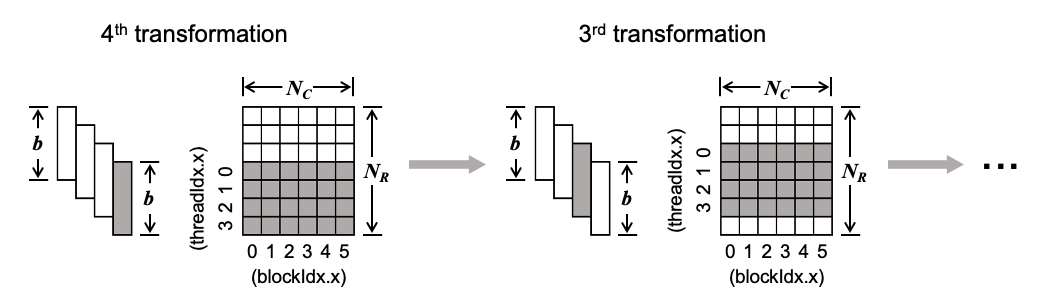}
\caption{Workflow of the Householder transformation CUDA kernel. The semi-bandwidth is $b=4$. Four Householder transformations are applied to the local eigenvector matrix with $N_R=7$ rows and $N_C=6$ columns. As indicated by the block index (blockIdx.x) and thread index (threadIdx.x), a CUDA block works on a column of the eigenvector matrix, while a thread within a block works on a single element. From the $n^\text{th}$ iteration to the $(n-1)^\text{th}$ iteration, the work set of the eigenvector matrix is shifted upward by one element.}
\label{fig:kernel}
\end{figure*}

An example of this procedure is given in Fig.~\ref{fig:kernel}, where four Householder transformations are applied to a local eigenvector matrix with $N_R=7$ rows and $N_C=6$ columns. The CUDA kernel is launched with six blocks and four threads, as indicated by the block id and thread id (both zero-based) in the figure. Throughout the execution of the kernel, block $i$ is responsible for the $(i+1)^\text{th}$ eigenvector. The four Householder transformations need to be applied in reverse order, i.e. from right to left. In the application of the rightmost transformation, threads 0, 1, 2, 3 in each block are responsible for the fourth, fifth, sixth, and seventh elements, respectively, of the eigenvector this block is responsible for. After computing Eq.~\ref{eq:kernel}, thread 3 immediately writes the seventh element back to the $\boldsymbol{X}_\text{local}$ array. Threads 1, 2, 3 take the updated fourth, fifth, and sixth elements from threads 0, 1, 2, respectively. Then thread 0 loads the third element of the eigenvector from $\boldsymbol{X}_\text{local}$, and all the threads are ready for the next transformation. It is obvious that all the transformations can be applied in exactly the same way. After the kernel finishes, the final $\boldsymbol{X}_\text{local}$ is copied back from GPU to CPU.

The semi-bandwidth $b=4$ in Fig.~\ref{fig:kernel} is chosen as an illustrative example only. On NVIDIA GPUs, a thread block is composed of warps. A warp is a group of 32 threads that execute the same instruction. When threads in a warp have to take different code paths, the so-called warp divergence problem occurs, often degrading the performance of the GPU code. In order to avoid warp divergence as much as possible, $b$ in our implementation is restricted to a multiple of 32, with 32, 64, and 128 being the recommended choices. The effect of $b$ on the performance of GPU-ELPA2 is examined in Sec.~\ref{subsec:results_params}. In step 2 (a) above, the parallel reduction across threads in the same warp can take advantage of the highly efficient CUDA warp-level primitives.

\section{Performance and Scalability}
\label{sec:results}
The performance of the GPU-accelerated ELPA2 solver is benchmarked on the Talos cluster at Max Planck Computing and Data Facility and the Summit supercomputer at Oak Ridge National Laboratory. Each node of Talos has two Intel Xeon Gold 6148 CPUs (40 cores in total) and two NVIDIA Tesla Volta V100 GPUs (each has 32 GB high-bandwidth memory, double precision peak 7.0 TFLOP/s, PCIe 32 GB/s interconnect). Benchmarks presented in Fig.~\ref{fig:all_talos} are performed on Talos. The ELPA code is compiled with the Intel 19.0.5 compiler suite, Intel MPI 2019.5, Intel MKL 2019.5, and CUDA 10.1. Each node of Summit has two IBM POWER9 CPUs (44 cores in total, of which 42 are for running applications) and six NVIDIA Tesla Volta V100 GPUs (each has 16 GB high-bandwidth memory, double precision peak 7.8 TFLOP/s, NVLink 50 GB/s interconnect). Benchmarks presented in Figs.~\ref{fig:all_summit}, \ref{fig:decomp3_summit}, \ref{fig:decomp5_summit}, \ref{fig:grid_summit} are performed on Summit. The ELPA code is compiled with the IBM XL 16.1.1 compiler suite, IBM Spectrum MPI 10.3, IBM ESSL 6.1, and CUDA 10.1.

In the benchmarks, the 2D MPI process grid is always chosen to be as close to square as possible. For example, a $14 \times 12$ grid is used for 168 MPI processes. NVIDIA MPS is enabled unless otherwise stated. In Figs.~\ref{fig:all_talos}, \ref{fig:all_summit}, \ref{fig:decomp3_summit}, \ref{fig:decomp5_summit}, and \ref{fig:summit_cori}, the block size of the 2D block-cyclic distribution is 16. For CPU-ELPA2, semi-bandwidths of 64 and 32 are used in the real and complex cases, respectively. For GPU-ELPA2, a semi-bandwidth of 32 is used in all cases. The effect of these parameters on the performance of the CPU version of ELPA has been discussed in Ref.~\cite{elpa_cook_2018}. Their effect on the GPU version is analyzed in Sec.~\ref{subsec:results_params} and Fig.~\ref{fig:grid_summit}.

\subsection{Overall Performance}
\label{subsec:results_overall}
\begin{figure*}[ht!]
\centering
\includegraphics[width=0.99\textwidth]{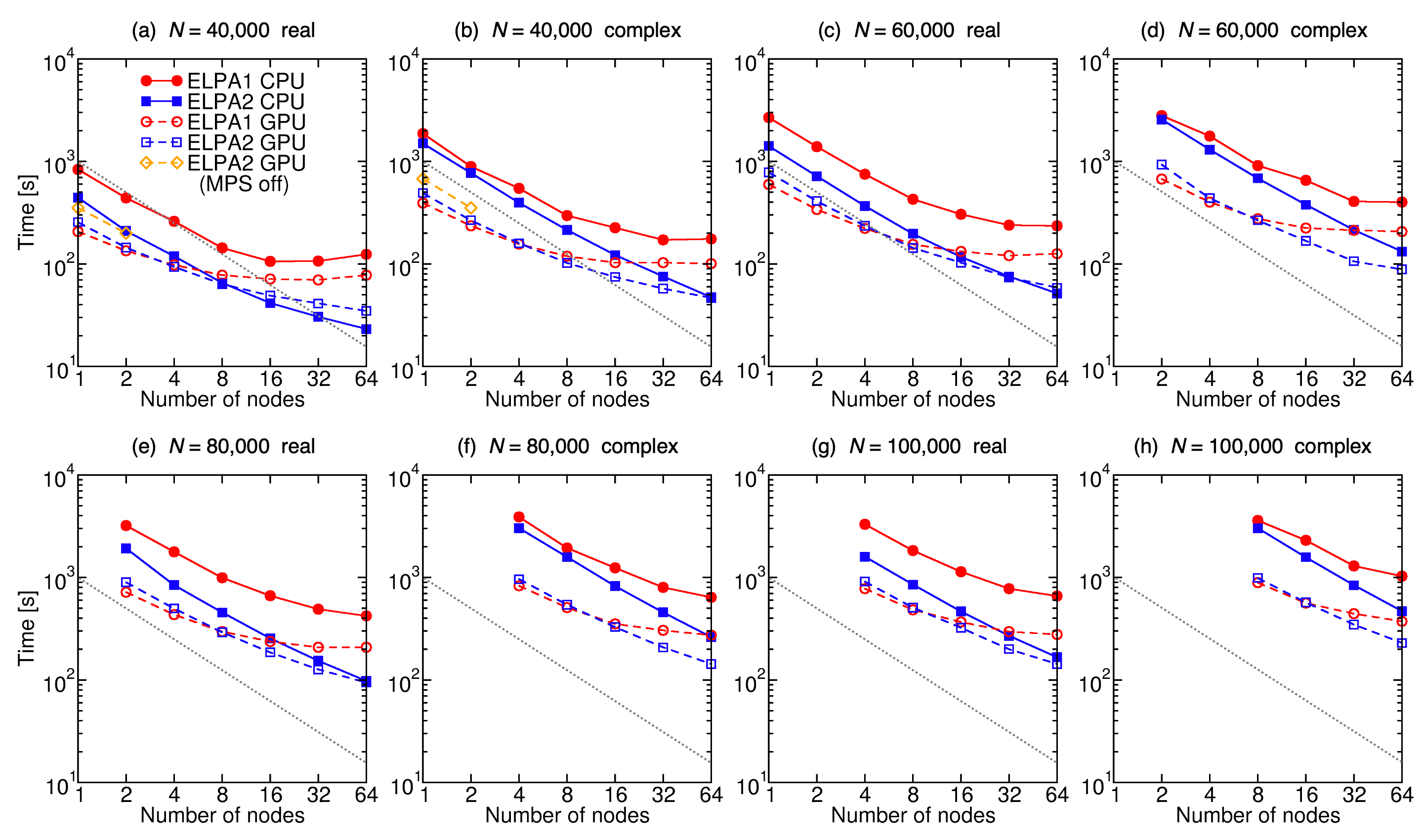}
\caption{Timings of CPU-ELPA1 (red solid), CPU-ELPA2 (blue solid), GPU-ELPA1 (MPS enabled, red dashed), GPU-ELPA2 (MPS enabled, blue dashed), and GPU-ELPA2 (``MPS off'', yellow dashed) for randomly generated, real symmetric and complex Hermitian matrices of size $N$ = 40,000, 60,000, 80,000, and 100,000. All eigenvalues and eigenvectors of the standard eigenproblem in Eq.~\ref{eq:evp} are computed. The gray dotted lines indicate ideal strong scaling. Tests are performed on the Talos cluster. The nodes are fully exploited by running 40 MPI processes (one MPI process per core, all cores per node) and two GPUs per node, i.e., 20 MPI processes accessing a single GPU, in both MPS-enabled and MPS-disabled (``MPS off'') cases.}
\label{fig:all_talos}
\end{figure*}

Fig.~\ref{fig:all_talos} shows the total time to solution of CPU-ELPA1, CPU-ELPA2, GPU-ELPA1, and GPU-ELPA2 on the Talos cluster. All eigenvalues and eigenvectors of randomly generated real and complex matrices of size $N$ = 40,000 to 100,000 are computed with up to 64 nodes. As already demonstrated in published benchmarks~\cite{elpa_marek_2014,elpa_gutheil_2014,elpa_cook_2018,elpa_auckenthaler_2013}, CPU-ELPA2 greatly outperforms CPU-ELPA1 in terms of performance and scalability. The performance difference between GPU-ELPA1 and GPU-ELPA2 is rather small. For small node counts, GPU-ELPA1 is marginally faster than GPU-ELPA2, which agrees with the published single-node tests~\cite{elpa_kus_2019b}. What has not been tested previously is the performance on multiple nodes. It turns out that GPU-ELPA2 becomes faster than GPU-ELPA1 as the node count increases. The crossover point depends on the dimension of the problem, e.g., four nodes in the $N$ = 40,000 real case, and 16 nodes in the $N$ = 100,000 complex case. When using 64 nodes, the speedup of GPU-ELPA2 over GPU-ELPA1, averaged over all matrix sizes in Fig.~\ref{fig:all_talos}, is 2.2x.

In Fig.~\ref{fig:all_talos}, the speedup enabled by the GPUs ranges from no speedup at all to 3.3x. Three general trends emerge: (1) For the same matrix size, the speedup becomes smaller as the node count increases. For small $N$, CPU-ELPA2 can even be faster than the GPU-accelerated solvers, thanks to the near-optimal strong scaling of CPU-ELPA2. (2) For the same number of nodes, the speedup becomes larger as the matrix size increases. (3) For the same node count and the same matrix size, the speedup is larger for complex matrices than for real matrices.

The benefit of using NVIDIA MPS is verified with $N$ = 40,000 on one and two Talos nodes, as shown in Fig.~\ref{fig:all_talos} (a) and (b). With 40 MPI processes and two GPUs per node, switching on MPS results in a speedup of over 20\%. When using MPS, CPU-GPU data transfers and GPU computations from different MPI processes can overlap with each other~\cite{mps}. This is confirmed by a profiling of the code using NVIDIA Visual Profiler.

\begin{figure*}[ht!]
\centering
\includegraphics[width=0.99\textwidth]{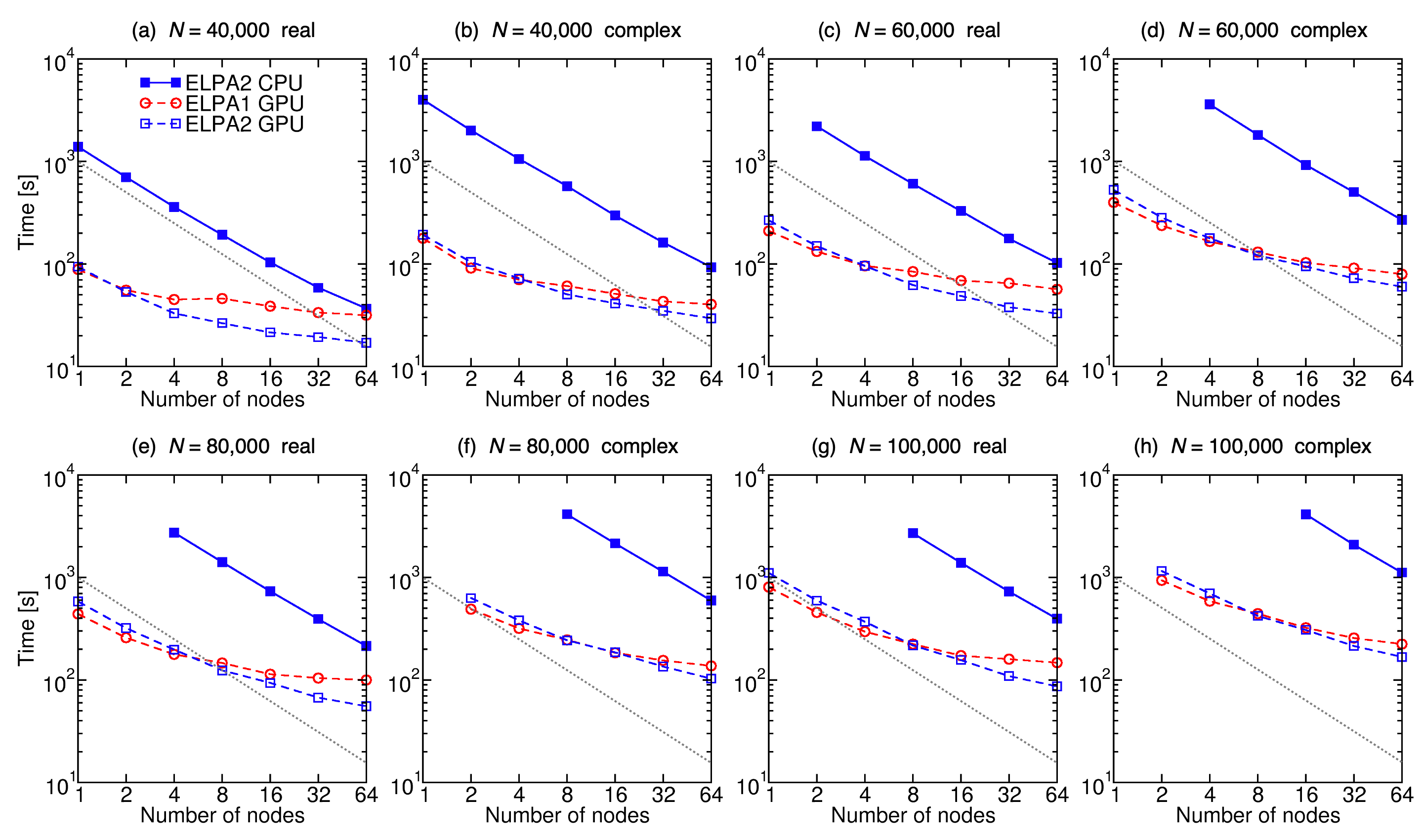}
\caption{Timings of CPU-ELPA2 (blue solid), GPU-ELPA1 (red dashed), and GPU-ELPA2 (blue dashed) for randomly generated, real symmetric and complex Hermitian matrices of size $N$ = 40,000, 60,000, 80,000, and 100,000. All eigenvalues and eigenvectors of the standard eigenproblem in Eq.~\ref{eq:evp} are computed. The gray dotted lines indicate ideal strong scaling. Tests are performed on the Summit supercomputer. The nodes are fully exploited by running 42 MPI processes (one MPI process per core, all cores per node) and six GPUs per node, i.e., seven MPI processes accessing a single GPU.}
\label{fig:all_summit}
\end{figure*}

The timing experiment in Fig.~\ref{fig:all_talos} is repeated on the Summit supercomputer. The results are shown in Fig.~\ref{fig:all_summit}. CPU-ELPA1 is omitted for simplicity. On Summit, the GPU-accelerated solvers GPU-ELPA1 and GPU-ELPA2 are always faster than the CPU-ELPA2 solver, with a maximum speedup of over 20x. The speedup of GPU-ELPA2 over CPU-ELPA2 remains 2.2x ($N$ = 40,000 real) to 6.7x ($N$ = 100,000 complex) even for 64 nodes. For the same matrix size and the same node count, the speedup on Summit appears greater than on Talos, which can be partially attributed to the difference in hardware. Summit has six GPUs per node, whereas Talos only has two GPUs per node. Data transfers on Summit take advantage of the NVLink technology~\cite{nvlink_foley_2017} for high-bandwidth interconnect between CPUs and GPUs. Besides, a high-performance CPU kernel for the tridiagonal-to-banded back-transformation, written in AVX512 instructions~\cite{elpa_kus_2019b}, is employed for the Intel Xeon Gold CPUs on Talos, rendering better performance of CPU-ELPA2 on Talos. Overall, we therefore observe that the absolute per-node timings of CPU-ELPA2 in Figs.~\ref{fig:all_talos} and \ref{fig:all_summit} are already lower on Talos than on Summit. This difference in the CPU-only results, which are the baseline of the reported speedups, probably exaggerates the comparison of GPU-ELPA2 and CPU-ELPA2 on Summit somewhat, relative to Talos. Nevertheless, the three trends summarized from the tests on Talos are still valid on Summit, that is, the GPU speedup is larger for (1) larger matrix size, (2) fewer nodes, and (3) solving a complex problem instead of a real one. On both computers, the strong scaling of the GPU solvers is never as good as that of the CPU solvers. This can be explained by the workload assigned to the individual nodes. When using a large number of nodes or solving a small matrix, the workload on each node becomes so little that the many GPUs cannot be saturated, and the cost of CPU-GPU communications cannot be amortized. In contrast, when solving a large matrix or using a small number of nodes, a large amount of local work is offloaded to the GPUs, resulting in a significant speedup.

\begin{figure*}[ht!]
\centering
\includegraphics[width=0.58\textwidth]{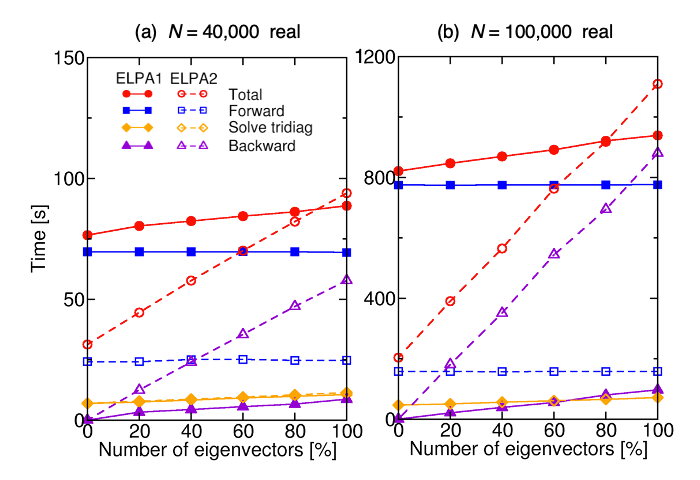}
\caption{Timings of GPU-ELPA1 (solid) and GPU-ELPA2 (dashed) as a function of the number of eigenvectors computed, for randomly generated, real symmetric matrices of size $N$ = 40,000 and 100,000. Tests are performed on one node of the Summit supercomputer. Shown are the total timing (red circles), the timings of the forward tridiagonalization (blue squares), the solution of the tridiagonal problem (yellow diamonds), and the back-transformation of the eigenvectors (violet triangles).}
\label{fig:decomp3_summit}
\end{figure*}

Note that in Figs.~\ref{fig:all_talos} and \ref{fig:all_summit}, all eigenvalues and eigenvectors are computed. In applications such as materials simulations based on KS-DFT, only a portion of the eigenspectrum, e.g., typically 20\% to 60\% for LCAO basis sets, is of interest. In this case, the advantage of ELPA2 over ELPA1 should be more significant, as the computational complexity of the back-transformation is proportional to the number of eigenvectors to be calculated. This is demonstrated in Fig.~\ref{fig:decomp3_summit}, where the total timings (red circles) of the GPU-ELPA1 (solid) and GPU-ELPA2 (dashed) are decomposed into the forward tridiagonalization (blue squares), the solution of the tridiagonal problem (yellow diamonds), and the backward transformation (violet triangles). Two test cases, namely $N$ = 40,000 real and $N$ = 100,000 real, are shown as examples for a single node on Summit, i.e., the worst-case scenario for ELPA2. The two-stage tridiagonalization in ELPA2 is always faster than the one-stage tridiagonalization in ELPA1 by a factor of approximately three. The back-transformation accounts for a small fraction of the total time in ELPA1, but is the most time-consuming part in ELPA2 when all the eigenvectors are computed. When computing fewer eigenvectors, the burden of the two-stage back-transformation in ELPA2 can be greatly alleviated, making ELPA2 more favorable than ELPA1.

Given that the optimal performance may be achieved with GPU-ELPA1, GPU-ELPA2, or CPU-ELPA2, depending on the specifics of the problem and the architecture, we highlight the auto-tuning feature in the ELPA library~\cite{elpa_kus_2019b}. When ELPA is called repeatedly, like e.g. in a self-consistent KS-DFT calculation, this auto-tuning feature automatically iterates over possible combinations of solvers and runtime parameters. The best solver for a given problem can be identified and utilized without any additional input from the user.

\subsection{Performance of Individual Computational Steps}
\label{subsec:results_steps}
\begin{figure*}[ht!]
\centering
\includegraphics[width=0.58\textwidth]{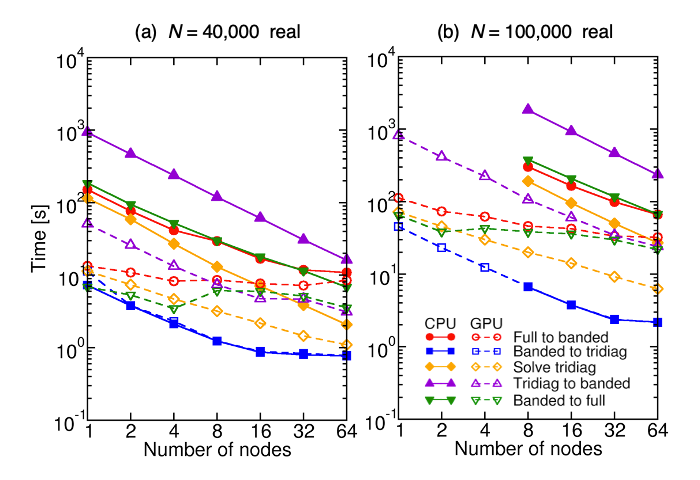}
\caption{Timing decomposition of CPU-ELPA2 (solid) and GPU-ELPA2 (dashed) for randomly generated, real symmetric matrices of size $N$ = 40,000 and 100,000. Tests are performed on the Summit supercomputer. Node utilization is identical to that in Fig.~\ref{fig:all_summit}. All eigenvalues and eigenvectors are computed. Shown are the timings of the full-to-banded transformation (red circles), the banded-to-tridiagonal transformation (blue squares), the solution of the tridiagonal problem (yellow diamonds), the tridiagonal-to-banded back-transformation (violet up triangles), and the banded-to-full back-transformation (green down triangles).}
\label{fig:decomp5_summit}
\end{figure*}

In Fig.~\ref{fig:decomp5_summit}, we further decompose the timings of CPU-ELPA2 and GPU-ELPA2 into the five computational steps defined in Eq.~\ref{eq:2stage}. Again, the $N$ = 40,000 real and $N$ = 100,000 real tests are shown as examples. Steps that have been GPU-accelerated display an excellent speedup, namely 5.8x, 6.1x, 17.1x, and 11.3x (averaged over all data points in Fig.~\ref{fig:all_summit}) for the full-to-banded transformation (Eq.~\ref{eq:2stage_fwd1}), the solution of the tridiagonal problem (Eq.~\ref{eq:2stage_solve}), the tridiagonal-to-banded back-transformation (Eq.~\ref{eq:2stage_bkwd1}), and the banded-to-full back-transformation (Eq.~\ref{eq:2stage_bkwd2}), respectively. The banded-to-tridiagonal transformation step (Eq.~\ref{eq:2stage_fwd2}) is not yet GPU-accelerated, as it never stands as a bottleneck. The tridiagonal-to-banded back-transformation, which uses the newly developed CUDA kernel described in Sec.~\ref{subsec:gpu_step4}, shows a strong scaling that is close to ideal. The scaling of the full-to-banded transformation and the banded-to-full back-transformation is not as good as the other steps. These two steps limit the overall parallel efficiency of GPU-ELPA2, therefore they would be the first target for further algorithmic and technical optimization.

\subsection{Runtime Parameters}
\label{subsec:results_params}
\begin{figure*}[ht!]
\centering
\includegraphics[width=0.3\textwidth]{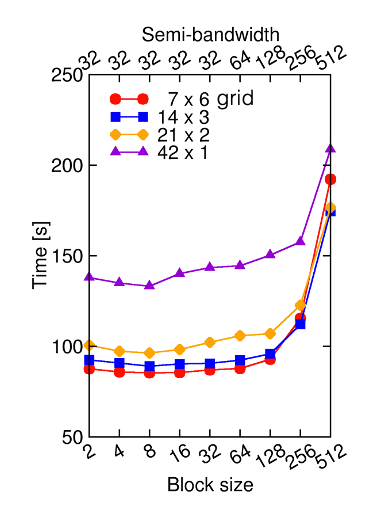}
\caption{Timings of GPU-ELPA2 for randomly generated, real symmetric matrices of size $N$ = 40,000 as a function of the block size, the semi-bandwidth, and the layout of the 2D MPI process grid. Tests are performed on one node of the Summit supercomputer. All eigenvalues and eigenvectors of the standard eigenproblem in Eq.~\ref{eq:evp} are computed. Shown are the timings obtained with $7 \times 6$ (red circles), $14 \times 3$ (blue squares), $21 \times 2$ (yellow diamonds), and $42 \times 1$ (violet up triangles) process grids.}
\label{fig:grid_summit}
\end{figure*}

Fig.~\ref{fig:grid_summit} compares the performance of GPU-ELPA2 on one node of Summit with different MPI process grids, namely $7 \times 6$, $14 \times 3$, $21 \times 2$, and $42 \times 1$. The $7 \times 6$ grid and the $42 \times 1$ grid yield the best performance and the worst performance, respectively. This matches the finding in Ref.~\cite{elpa_cook_2018} that CPU-ELPA2 prefers a square-like grid. MPI communications in ELPA2 take place along either the row direction or the column direction of the process grid. Communications in both directions are well balanced within a square grid. In contrast, when using a long rectangular grid, communications along one direction may become more expensive than those along the other direction.

Fig.~\ref{fig:grid_summit} also shows the effect of block size and semi-bandwidth on the performance of GPU-ELPA2. As explained in Sec.~\ref{subsec:gpu_step4}, the semi-bandwidth is at least 32 in order to minimize warp divergence in the CUDA kernel. When the block size is larger than 32, GPU-ELPA2 by default uses a semi-bandwidth equal to the block size. For a fixed process grid, the performance of GPU-ELPA2 is near optimal for a wide range of block sizes between 2 and 128, but deteriorates as the block size exceeds 128. This is mainly due to the computational steps that have not been ported to GPUs, notably the banded-to-tridiagonal transformation of which the computational cost increases proportionally with respect to the semi-bandwidth.

The performance of GPU-ELPA2 appears to be insensitive to the CPU-GPU affinity. On each socket of a Summit node, there are 21 CPU cores and three GPUs. The tests in Fig.~\ref{fig:grid_summit} are repeated with two layouts of CPU-GPU affinity: (1) Block: MPI processes 0--6 use GPU 0, MPI processes 7--13 use GPU 1, MPI processes 14--20 use GPU 2. (2) Cyclic: MPI processes 0, 3, 6, ..., 18 use GPU 0, MPI processes 1, 4, 7, ..., 19 use GPU 1, MPI processes 2, 5, 8, ..., 20 use GPU 2. The cyclic layout is marginally faster by 8\% averaged over the cases in Fig.~\ref{fig:grid_summit}.

\section{Application in Materials Calculations}
\label{sec:app}
\begin{figure*}[htb!]
\centering
\includegraphics[width=0.7\textwidth]{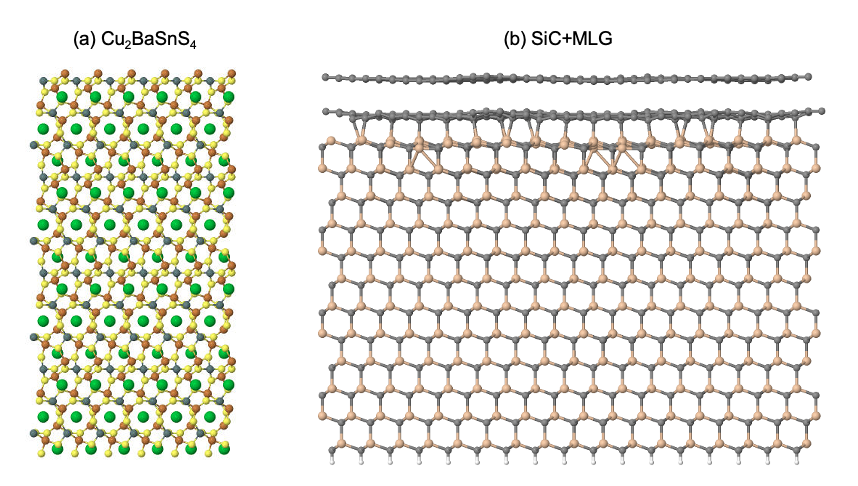}
\caption{Atomic structures of (a) Cu$_2$BaSnS$_4$ and (b) graphene on SiC.}
\label{fig:geometry}
\end{figure*}

Finally, we demonstrate the efficiency of GPU-ELPA2 in actual materials simulations. Two atomic structure models are selected as test systems, namely Cu$_2$BaSnS$_4$ and graphene on SiC as shown in Fig.~\ref{fig:geometry} (a) and (b), respectively. Cu$_2$BaSnS$_4$ is a semiconductor that was designed as a potential photovoltaic absorber material~\cite{cbts_shin_2016,cbts_shin_2017}. In this class of materials, large supercells can be of importance to understand, for example, the impact of dynamical properties, of defects, or of deliberately introduced dopant materials. Graphene on SiC features a monolayer graphene of very high crystalline quality that can be obtained as a thermodynamic equilibrium phase~\cite{sic_nemec_2013} on a SiC surface. The combined system has a very large $(6 \sqrt{3} \times 6 \sqrt{3})$-R30$^{\circ}$ crystallographic unit cell. Calculations of this system have, for instance, been used to gain insights into subsurface structure perturbations by comparing to experimentally obtained signals from contact-resonance atomic force microscopy~\cite{sic_tu_2016}.

The crystallographic unit cell of Cu$_2$BaSnS$_4$ contains 24 atoms. For this number of atoms, simulations based on semi-local DFT or a higher level of theory are affordable even on desktop computers. However, the number of atoms and the associated computational cost increase rapidly with respect to the size of the supercell. To test the performance of the GPU-ELPA2 solver, KS-DFT calculations with the PBE exchange-correlation functional are carried out using the all-electron full-potential FHI-aims code~\cite{fhiaims_blum_2009,gpu_huhn_2020} for $2 \times 2 \times 2$, $3 \times 3 \times 3$, $4 \times 4 \times 4$, and $5 \times 5 \times 5$ supercell models of Cu$_2$BaSnS$_4$. The ELPA library is connected to FHI-aims through the ELSI interface~\cite{elsi_yu_2018,elsi_yu_2020}. In these calculations, the default ``light'' numerical settings of FHI-aims and a $1 \times 1 \times 1$ $\boldsymbol{k}$-point grid are used. The number of atoms ranges from 192 to 3,000, and the number of basis functions ranges from 5,136 to 80,250.

\begin{figure*}[htb!]
\centering
\includegraphics[width=0.3\textwidth]{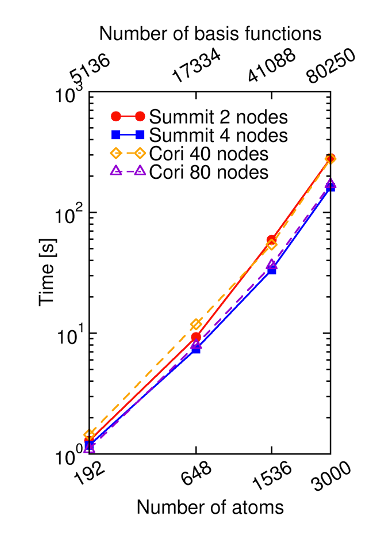}
\caption{Performance comparison of CPU-ELPA2 and GPU-ELPA2 in KS-DFT calculations of Cu$_2$BaSnS$_4$ models, as a function of the number of atoms. All the eigenvalues and 69.2\% of the eigenvectors of the generalized eigenproblem in Eq.~\ref{eq:gevp} are computed. Computational steps included in the timings are Eqs.~\ref{eq:evp}, \ref{eq:gevp2evp}, and \ref{eq:evp2gevp}, but not Eq.~\ref{eq:cholesky}. CPU-ELPA2 and GPU-ELPA2 results are obtained by running the FHI-aims code on the Cori supercomputer (Haswell partition) and the Summit supercomputer, respectively. Red circles: GPU-ELPA2 on two Summit nodes. Blue squares: GPU-ELPA2 on four Summit nodes. Yellow diamonds: CPU-ELPA2 on 40 Cori-Haswell nodes. Violet triangles: CPU-ELPA2 on 80 Cori-Haswell nodes.}
\label{fig:summit_cori}
\end{figure*}

The timings are shown in Fig.~\ref{fig:summit_cori}, with CPU-ELPA2 timings obtained on the Cori supercomputer (Haswell partition) included for comparison. The theoretical peak performances of one Summit node and one Cori-Haswell node are 42 TFLOP/s and 1.2 TFLOP/s, respectively. In this particular test case, the performance of GPU-ELPA2 on two (four) Summit nodes is comparable to that of CPU-ELPA2 on 40 (80) Cori-Haswell nodes, though a higher fraction of the peak performance is achieved on Cori-Haswell.

Table~\ref{tab:summit_cori} shows the timing of one self-consistent field (SCF) iteration in the largest calculation in Fig.~\ref{fig:summit_cori}, i.e., the $5 \times 5 \times 5$ Cu$_2$BaSnS$_4$ supercell with 3,000 atoms. The total timing is decomposed into the computation of the electrostatic potential, the numerical integration of the Hamiltonian matrix, the solution of the generalized eigenproblem (ELPA2), and the computation of the electron density. When running FHI-aims on Summit, GPU acceleration is enabled in the Hamiltonian integration, the density computation, and the eigensolver. The electrostatic potential is computed on CPUs, as the code has not been ported to GPUs. On Cori, the time spent on CPU-ELPA2 amounts to more than 85\% of the total time. In contrast, GPU-ELPA2 on Summit takes only about 45\% of the total time, to a fair extent eliminating the key computational bottleneck in large simulations based on semi-local KS-DFT.
\begin{table*}[htb!]
\centering
\caption{Performance comparison of GPU-accelerated and CPU-only KS-DFT calculations of the Cu$_2$BaSnS$_4$ model (CBTS) and the graphene on SiC model (SiC+G) in Fig.~\ref{fig:geometry}. Timings are obtained by running the FHI-aims code on N$_\text{node}$ nodes of the Summit supercomputer and the Cori supercomputer (Haswell partition). The total time of one SCF iteration is decomposed into the computation of the electrostatic potential, the numerical integration of the Hamiltonian matrix, the solution of the generalized eigenproblem (ELPA2), and the computation of the electron density. Complete input and output of the FHI-aims calculations are provided online~\cite{raw_data}.
\newline * The electrostatic potential is computed on CPUs.}
\scriptsize
\begin{tabular}{c c c c c c c c}
\hline
\hline
\multirow{2}{*}{system} & \multirow{2}{*}{computer} & \multirow{2}{*}{N$_\text{node}$} & \multicolumn{5}{c}{time [s]} \\
\cline{4-8}
& & & potential & Hamiltonian & eigenproblem & density & total \\
\hline
              & Summit &  2 & 246.1* & 21.3 & 280.7 & 50.8 & 598.9 \\
CBTS          & Summit &  4 & 123.4* & 12.8 & 161.4 & 28.2 & 325.8 \\
(3,000 atoms) &   Cori & 40 &  15.1  &  3.7 & 277.8 & 23.6 & 320.2 \\
              &   Cori & 80 &  10.3  &  1.9 & 170.7 & 16.1 & 199.0 \\
\hline
              & Summit &  2 & 164.7* & 15.0 &  88.5 & 28.1 & 296.3 \\
SiC+G         & Summit &  4 &  82.6* &  7.5 &  58.4 & 15.7 & 164.2 \\
(3,376 atoms) &   Cori & 40 &  10.8  &  2.8 &  81.4 &  8.6 & 103.6 \\
              &   Cori & 80 &   6.6  &  1.4 &  60.9 &  6.8 &  75.7 \\
\hline
\hline
\end{tabular}
\label{tab:summit_cori}
\end{table*}

The structure in Fig.~\ref{fig:geometry} (b) is a strain-free $(6 \sqrt{3} \times 6 \sqrt{3})$-R30$^{\circ}$ interface model of graphene on SiC substrate. The 3,376-atom model contains 12 bilayers of SiC. The carbon atoms in the bottom layer are saturated by hydrogen atoms. On top of the substrate, there are two layers of carbon atoms, corresponding to a carbon buffer layer and a graphene monolayer. As done in the Cu$_2$BaSnS$_4$ case, we report in Table~\ref{tab:summit_cori} the timing decomposition of one SCF iteration of KS-DFT calculations, performed with the FHI-aims code on the Summit and Cori supercomputers. The performance of GPU-ELPA2 on two and four Summit nodes generally matches the performance of CPU-ELPA2 on 40 and 80 Cori-Haswell nodes, respectively. GPU-ELPA2 on Summit consumes a much smaller fraction of the total computation time compared to CPU-ELPA2 on Cori. We note again that at the time of writing, the computation of the electrostatic potential in FHI-aims is not able to use GPUs, hence its relatively poor performance on Summit. This also implies that in this particular case, the formally $\mathcal{O}(N^3)$ scaling eigenproblem is no longer the top computational bottleneck, which we consider as a success of the GPU-ELPA2 solver.

The Cu$_2$BaSnS$_4$ and graphene on SiC models are two examples of many similar areas where large scale simulations can connect experiment to underlying computationally derived, atomistic insights. In the benchmark calculations reported in this section, the difference in the total energies computed with CPU-ELPA2 and GPU-ELPA2 lies below $10^{-6}$ eV, or $10^{-9}$ eV/atom. The availability of the GPU-ELPA2 solver accelerates semi-local KS-DFT calculations comprising at least thousands of atoms without sacrificing any accuracy. It is thus expected to strengthen the field of computational physics and materials science significantly.

\section{Conclusions}
\label{sec:conclude}
In this paper, we report our GPU-oriented optimizations of the two-stage tridiagonalization eigensolver ELPA2 in the ELPA library. The local BLAS operations in ELPA2 are offloaded to GPUs via the cuBLAS library. The tridiagonal-to-banded back-transformation of eigenvectors, which cannot be easily written as BLAS operations, is GPU-accelerated by a CUDA implementation. The overall performance of the GPU-accelerated ELPA2 solver is promising. It delivers a significant performance boost over the CPU-only version of ELPA2, as demonstrated by benchmarks on the Talos and Summit computers. Owing to the advanced two-stage tridiagonalization algorithm, the parallel scaling of the GPU-ELPA2 solver is superior to that of the GPU-ELPA1 solver. Based on an analysis of the individual computational steps in ELPA2, we identify the full-to-banded transformation and banded-to-full back-transformation steps as the next target for future optimization, as their strong scaling is not yet optimal. Nevertheless, the GPU-ELPA2 solver in its current form already unlocks a significant potential of GPU computations, as exemplified by KS-DFT based simulations of thousands of atoms.

The code developed in this paper is available in the ELPA library since its 2020.05.001 release. Information about obtaining, installing, and using ELPA may be found at \url{https://elpa.mpcdf.mpg.de} .

\section*{Acknowledgments}
\label{sec:thanks}
This research was supported by the National Science Foundation (NSF) under Award No. 1450280, and was partly supported by the U.S. Department of Energy, Office of Science, Office of Basic Energy Sciences, Materials Sciences and Engineering Division. Yu was supported by a fellowship from the Molecular Sciences Software Institute under NSF Award No. 1547580. This research used resources of the Oak Ridge Leadership Computing Facility at the Oak Ridge National Laboratory, which is supported by the Office of Science of the U.S. Department of Energy under Contract No. DE-AC05-00OR22725. This research also used resources of the National Energy Research Scientific Computing Center (NERSC), a U.S. Department of Energy Office of Science User Facility operated under Contract No. DE-AC02-05CH11231. We acknowledge support from NVIDIA Corporation through the donation of three GPUs used for development purposes in our research. We thank Dr. Christoph Angerer and Scott Biersdorff from NVIDIA, Dr. Bj\"{o}rn Lange, Dr. William Huhn, and Dr. Wenhui Mi from Duke University, Dr. Jack Deslippe, Dr. Mauro Del Ben, and Dr. Charlene Yang from NERSC, Dr. Eduardo D'Azevedo from Oak Ridge Leadership Computing Facility (OLCF), Dr. Markus Rampp from Max Planck Computing and Data Facility, Garching, Germany, and Dr. Christian Carbogno and Dr. Matthias Scheffler from Fritz Haber Institute of the Max Planck Society, Berlin, Germany for fruitful discussions. Part of the optimization presented in this paper was the outcome of the 2019 OLCF GPU Hackathon. We gratefully acknowledge the organizer Dr. Thomas Papatheodore (OLCF) and mentors Brent Leback (NVIDIA) and Dr. Markus Eisenbach (OLCF) for their advice during this event. We thank the reviewers for their constructive comments.

\bibliographystyle{elsarticle-num}
\bibliography{elpa2}

\begin{thebibliography}{10}
\expandafter\ifx\csname url\endcsname\relax
  \def\url#1{\texttt{#1}}\fi
\expandafter\ifx\csname urlprefix\endcsname\relax\def\urlprefix{URL }\fi
\expandafter\ifx\csname href\endcsname\relax
  \def\href#1#2{#2} \def\path#1{#1}\fi

\bibitem{dft_hohenberg_1964}
P.~Hohenberg, W.~Kohn, Inhomogeneous electron gas, Physical Review 136 (1964)
  B864--B871.

\bibitem{dft_kohn_1965}
W.~Kohn, L.~J. Sham, Self-consistent equations including exchange and
  correlation effects, Physical Review 140 (1965) A1133--A1138.

\bibitem{conquest_nakata_2020}
A.~Nakata, J.~S. Baker, S.~Y. Mujahed, J.~T.~L. Poulton, S.~Arapan, J.~Lin,
  Z.~Raza, S.~Yadav, L.~Truflandier, T.~Miyazaki, D.~R. Bowler, Large scale and
  linear scaling {DFT} with the {CONQUEST} code, The Journal of Chemical
  Physics 152 (2020) 164112.

\bibitem{fhiaims_blum_2009}
V.~Blum, R.~Gehrke, F.~Hanke, P.~Havu, V.~Havu, X.~Ren, K.~Reuter,
  M.~Scheffler, Ab initio molecular simulations with numeric atom-centered
  orbitals, Computer Physics Communications 180 (2009) 2175--2196.

\bibitem{onetep_prentice_2020}
J.~C.~A. Prentice, J.~Aarons, J.~C. Womack, A.~E.~A. Allen, L.~Andrinopoulos,
  L.~Anton, R.~A. Bell, A.~Bhandari, G.~A. Bramley, R.~J. Charlton, R.~J.
  Clements, D.~J. Cole, G.~Constantinescu, F.~Corsetti, S.~M.~M. Dubois,
  K.~K.~B. Duff, J.~M. Escart\'{i}n, A.~Greco, Q.~Hill, L.~P. Lee, E.~Linscott,
  D.~D. O'Regan, M.~J.~S. Phipps, L.~E. Ratcliff, \'{A}lvaro Ruiz~Serrano,
  E.~W. Tait, G.~Teobaldi, V.~Vitale, N.~Yeung, T.~J. Zuehlsdorff, J.~Dziedzic,
  P.~D. Haynes, N.~D.~M. Hine, A.~A. Mostofi, M.~C. Payne, C.-K. Skylaris, The
  {ONETEP} linear-scaling density functional theory program, The Journal of
  Chemical Physics 152 (2020) 174111.

\bibitem{siesta_garcia_2020}
A.~Garc\'{i}a, N.~Papior, A.~Akhtar, E.~Artacho, V.~Blum, E.~Bosoni,
  P.~Brandimarte, M.~Brandbyge, J.~I. Cerd\'{a}, F.~Corsetti, R.~Cuadrado,
  V.~Dikan, J.~Ferrer, J.~Gale, P.~Garc\'{i}a-Fern\'{a}ndez, V.~M.
  Garc\'{i}a-Su\'{a}rez, S.~Garc\'{i}a, G.~Huhs, S.~Illera, R.~Koryt\'{a}r,
  P.~Koval, I.~Lebedeva, L.~Lin, P.~L\'{o}pez-Tarifa, S.~G. Mayo, S.~Mohr,
  P.~Ordej\'{o}n, A.~Postnikov, Y.~Pouillon, M.~Pruneda, R.~Robles,
  D.~S\'{a}nchez-Portal, J.~M. Soler, R.~Ullah, V.~W.-z. Yu, J.~Junquera, The
  {Siesta} method: Recent developments and applications, The Journal of
  Chemical Physics 152 (2020) 204108.

\bibitem{nersc_2014}
2014 {NERSC} workload analysis,
  \url{https://portal.nersc.gov/project/mpccc/baustin/NERSC\_2014\_Workload\_Analysis\_v1.1.pdf}
  (Accessed: 2020-09-07).

\bibitem{nersc_2018}
{NERSC}-10 workload analysis (data from 2018),
  \url{https://portal.nersc.gov/project/m888/nersc10/workload/N10\_Workload\_Analysis.latest.pdf}
  (Accessed: 2020-09-07).

\bibitem{elsi_yu_2018}
V.~W.-z. Yu, F.~Corsetti, A.~Garc\'{i}a, W.~P. Huhn, M.~Jacquelin, W.~Jia,
  B.~Lange, L.~Lin, J.~Lu, W.~Mi, A.~Seifitokaldani, \'{A}lvaro
  V\'{a}zquez-Mayagoitia, C.~Yang, H.~Yang, V.~Blum, {ELSI}: A unified software
  interface for {Kohn-Sham} electronic structure solvers, Computer Physics
  Communications 222 (2018) 267--285.

\bibitem{elsi_yu_2020}
V.~W.-z. Yu, C.~Campos, W.~Dawson, A.~Garc\'{i}a, V.~Havu, B.~Hourahine, W.~P.
  Huhn, M.~Jacquelin, W.~Jia, M.~Ke\c{c}eli, R.~Laasner, Y.~Li, L.~Lin, J.~Lu,
  J.~Moussa, J.~E. Roman, \'{A}lvaro V\'{a}zquez-Mayagoitia, C.~Yang, V.~Blum,
  {ELSI} -- an open infrastructure for electronic structure solvers, Computer
  Physics Communications 256 (2020) 107459.

\bibitem{numerical_press_2007}
W.~H. Press, S.~A. Teukolsky, W.~T. Vetterling, B.~P. Flannery, Numerical
  recipes 3rd edition: The art of scientific computing, Cambridge university
  press, 2007.

\bibitem{matrix_golub_2013}
G.~H. Golub, C.~F. {van Loan}, Matrix Computations, Johns Hopkins Studies in
  the Mathematical Sciences, Johns Hopkins University Press, 2013.

\bibitem{eigenexa_imamura_2011}
T.~Imamura, S.~Yamada, M.~Machida, Development of a high-performance
  eigensolver on a peta-scale next-generation supercomputer system, Progress in
  Nuclear Science and Technology 2 (2011) 643--650.

\bibitem{2stage_lang_1993}
B.~Lang, A parallel algorithm for reducing symmetric banded matrices to
  tridiagonal form, SIAM Journal on Scientific Computing 14 (1993) 1320--1338.

\bibitem{2stage_bischof_1994}
C.~Bischof, X.~Sun, B.~Lang, Parallel tridiagonalization through two-step band
  reduction, in: Proceedings of {IEEE} Scalable High Performance Computing
  Conference, 1994, pp. 23--27.

\bibitem{elpa_auckenthaler_2011}
T.~Auckenthaler, V.~Blum, H.~J. Bungartz, T.~Huckle, R.~Johanni, L.~Kramer,
  B.~Lang, H.~Lederer, P.~R. Willems, Parallel solution of partial symmetric
  eigenvalue problems from electronic structure calculations, Parallel
  Computing 37 (2011) 783--794.

\bibitem{elpa_marek_2014}
A.~Marek, V.~Blum, R.~Johanni, V.~Havu, B.~Lang, T.~Auckenthaler, A.~Heinecke,
  H.~J. Bungartz, H.~Lederer, The {ELPA} library: Scalable parallel eigenvalue
  solutions for electronic structure theory and computational science, Journal
  of Physics: Condensed Matter 26 (2014) 213201.

\bibitem{davidson_davidson_1975}
E.~R. Davidson, The iterative calculation of a few of the lowest eigenvalues
  and corresponding eigenvectors of large real-symmetric matrices, Journal of
  Computational Physics 17 (1975) 87--94.

\bibitem{davidson_sleijpen_1996}
G.~L.~G. Sleijpen, H.~A. {van der Vorst}, A {Jacobi-Davidson} iteration method
  for linear eigenvalue problems, SIAM Journal on Matrix Analysis and
  Applications 17 (1996) 401--425.

\bibitem{iterative_payne_1992}
M.~C. Payne, M.~P. Teter, D.~C. Allan, T.~A. Arias, J.~D. Joannopoulos,
  Iterative minimization techniques for ab initio total-energy calculations:
  Molecular dynamics and conjugate gradients, Reviews of Modern Physics 64
  (1992) 1045.

\bibitem{iterative_kresse_1996}
G.~Kresse, J.~Furthm\"{u}ller, Efficient iterative schemes for ab initio
  total-energy calculations using a plane-wave basis set, Physical Review B 54
  (1996) 11169.

\bibitem{linear_goedecker_1999}
S.~Goedecker, Linear scaling electronic structure methods, Reviews of Modern
  Physics 71 (1999) 1085.

\bibitem{linear_bowler_2012}
D.~R. Bowler, T.~Miyazaki, Methods in electronic structure calculations,
  Reports on Progress in Physics 75 (2012) 036503.

\bibitem{linear_moussa_2019}
J.~Moussa, A.~Baczewski, Assessment of localized and randomized algorithms for
  electronic structure, Electronic Structure 1 (2019) 033001.

\bibitem{feast_polizzi_2009}
E.~Polizzi, Density-matrix-based algorithm for solving eigenvalue problems,
  Physical Review B 79 (2009) 115112.

\bibitem{pexsi_lin_2013}
L.~Lin, M.~Chen, C.~Yang, L.~He, Accelerating atomic orbital-based electronic
  structure calculation via pole expansion and selected inversion, Journal of
  Physics: Condensed Matter 25 (2013) 295501.

\bibitem{ntpoly_dawson_2018}
W.~Dawson, T.~Nakajima, Massively parallel sparse matrix function calculations
  with {NTPoly}, Computer Physics Communications 225 (2018) 154--165.

\bibitem{million_bowler_2010}
D.~R. Bowler, T.~Miyazaki, Calculations for millions of atoms with density
  functional theory: Linear scaling shows its potential, Journal of Physics:
  Condensed Matter 22 (2010) 074207.

\bibitem{million_vandevondele_2012}
J.~VandeVondele, U.~Bor\v{s}tnik, J.~Hutter, Linear scaling self-consistent
  field calculations with millions of atoms in the condensed phase, Journal of
  Chemical Theory and Computation 8 (2012) 3565--3573.

\bibitem{top500}
\url{https://top500.org} (Accessed: 2020-12-27).

\bibitem{cusolver}
\url{https://docs.nvidia.com/cuda/cusolver} (Accessed: 2020-09-07).

\bibitem{magma_tomov_2010}
S.~Tomov, J.~Dongarra, M.~Baboulin, Towards dense linear algebra for hybrid
  {GPU} accelerated manycore systems, Parallel Computing 36 (2010) 232--240.

\bibitem{magma_dongarra_2014}
J.~Dongarra, M.~Gates, A.~Haidar, J.~Kurzak, P.~Luszczek, S.~Tomov,
  I.~Yamazaki, Accelerating numerical dense linear algebra calculations with
  {GPUs}, in: Numerical computations with GPUs, Springer, 2014, pp. 3--28.

\bibitem{elpa}
\url{https://elpa.mpcdf.mpg.de} (Accessed: 2020-09-07).

\bibitem{abinit_gonze_2020}
X.~Gonze, B.~Amadon, G.~Antonius, F.~Arnardi, L.~Baguet, J.-M. Beuken,
  J.~Bieder, F.~Bottin, J.~Bouchet, E.~Bousquet, N.~Brouwer, F.~Bruneval,
  G.~Brunin, T.~Cavignac, J.-B. Charraud, W.~Chen, M.~C\^{o}t\'{e},
  S.~Cottenier, J.~Denier, G.~Geneste, P.~Ghosez, M.~Giantomassi, Y.~Gillet,
  O.~Gingras, D.~R. Hamann, G.~Hautier, X.~He, N.~Helbig, N.~Holzwarth, Y.~Jia,
  F.~Jollet, W.~Lafargue-Dit-Hauret, K.~Lejaeghere, M.~A.~L. Marques,
  A.~Martin, C.~Martins, H.~P.~C. Miranda, F.~Naccarato, K.~Persson,
  G.~Petretto, V.~Planes, Y.~Pouillon, S.~Prokhorenko, F.~Ricci, G.-M.
  Rignanese, A.~H. Romero, M.~M. Schmitt, M.~Torrent, M.~J. {van Setten},
  B.~{Van Troeye}, M.~J. Verstraete, G.~Z\'{e}rah, J.~W. Zwanziger, The
  {ABINIT} project: Impact, environment and recent developments, Computer
  Physics Communications 248 (2020) 107042.

\bibitem{berkeleygw_deslippe_2012}
J.~Deslippe, G.~Samsonidze, D.~A. Strubbe, M.~Jain, M.~L. Cohen, S.~G. Louie,
  {BerkeleyGW}: A massively parallel computer package for the calculation of
  the quasiparticle and optical properties of materials and nanostructures,
  Computer Physics Communications 183 (2012) 1269--1289.

\bibitem{cp2k_kuhne_2020}
T.~D. K\"{u}hne, M.~Iannuzzi, M.~{Del Ben}, V.~V. Rybkin, P.~Seewald, F.~Stein,
  T.~Laino, R.~Z. Khaliullin, O.~Sch\"{u}tt, F.~Schiffmann, D.~Golze,
  J.~Wilhelm, S.~Chulkov, M.~H. Bani-Hashemian, V.~Weber, U.~Bor\v{s}tnik,
  M.~Taillefumier, A.~S. Jakobovits, A.~Lazzaro, H.~Pabst, T.~M\"{u}ller,
  R.~Schade, M.~Guidon, S.~Andermatt, N.~Holmberg, G.~K. Schenter, A.~Hehn,
  A.~Bussy, F.~Belleflamme, G.~Tabacchi, A.~Gl\"{o}{\ss}, M.~Lass, I.~Bethune,
  C.~J. Mundy, C.~Plessl, M.~Watkins, J.~VandeVondele, M.~Krack, J.~Hutter,
  {CP2K}: An electronic structure and molecular dynamics software package --
  {Quickstep}: Efficient and accurate electronic structure calculations, The
  Journal of Chemical Physics 152 (2020) 194103.

\bibitem{cpmd_kloffel_2021}
T.~Kl\"{o}ffel, G.~Mathias, B.~Meyer, Integrating state of the art compute,
  communication, and autotuning strategies to multiply the performance of ab
  initio molecular dynamics on massively parallel multi-core supercomputers,
  Computer Physics Communications 260 (2021) 107745.

\bibitem{dftb_hourahine_2020}
B.~Hourahine, B.~Aradi, V.~Blum, F.~Bonaf\'{e}, A.~Buccheri, C.~Camacho,
  C.~Cevallos, M.~Y. Deshaye, T.~Dumitric\u{a}, A.~Dominguez, S.~Ehlert,
  M.~Elstner, T.~{van der Heide}, J.~Hermann, S.~Irle, J.~J. Kranz,
  C.~K\"{o}hler, T.~Kowalczyk, T.~Kuba\v{r}, I.~S. Lee, V.~Lutsker, R.~J.
  Maurer, S.~K. Min, I.~Mitchell, C.~Negre, T.~A. Niehaus, A.~M.~N. Niklasson,
  A.~J. Page, A.~Pecchia, G.~Penazzi, M.~P. Persson, J.~\v{R}ez\'{a}\v{c},
  C.~G. S\'{a}nchez, M.~Sternberg, M.~St\"{o}hr, F.~Stuckenberg, A.~Tkatchenko,
  V.~W.-z. Yu, T.~Frauenheim, {DFTB+}, a software package for efficient
  approximate density functional theory based atomistic simulations, The
  Journal of Chemical Physics 152 (2020) 124101.

\bibitem{gpaw_enkovaara_2010}
J.~Enkovaara, C.~Rostgaard, J.~J. Mortensen, J.~Chen, M.~Du{\l}ak, L.~Ferrighi,
  J.~Gavnholt, C.~Glinsvad, V.~Haikola, H.~A. Hansen, H.~H. Kristoffersen,
  M.~Kuisma, A.~H. Larsen, L.~Lehtovaara, M.~Ljungberg, O.~Lopez-Acevedo, P.~G.
  Moses, J.~Ojanen, T.~Olsen, V.~Petzold, N.~A. Romero, J.~Stausholm-M{\o}ller,
  M.~Strange, G.~A. Tritsaris, M.~Vanin, M.~Walter, B.~Hammer, H.~H\"{a}kkinen,
  G.~K.~H. Madsen, R.~M. Nieminen, J.~K. N{\o}rskov, M.~Puska, T.~T. Rantala,
  J.~Schi{\o}tz, K.~S. Thygesen, K.~W. Jacobsen, Electronic structure
  calculations with {GPAW}: A real-space implementation of the projector
  augmented-wave method, Journal of Physics: Condensed Matter 22 (2010) 253202.

\bibitem{nwchem_apra_2020}
E.~Apr\`{a}, E.~J. Bylaska, W.~A. {de Jong}, N.~Govind, K.~Kowalski, T.~P.
  Straatsma, M.~Valiev, H.~J.~J. {van Dam}, Y.~Alexeev, J.~Anchell,
  V.~Anisimov, F.~W. Aquino, R.~Atta-Fynn, J.~Autschbach, N.~P. Bauman, J.~C.
  Becca, D.~E. Bernholdt, K.~Bhaskaran-Nair, S.~Bogatko, P.~Borowski,
  J.~Boschen, J.~Brabec, A.~Bruner, E.~Cau\"{e}t, Y.~Chen, G.~N. Chuev, C.~J.
  Cramer, J.~Daily, M.~J.~O. Deegan, T.~H. Dunning, M.~Dupuis, K.~G. Dyall,
  G.~I. Fann, S.~A. Fischer, A.~Fonari, H.~Fr\"{u}chtl, L.~Gagliardi, J.~Garza,
  N.~Gawande, S.~Ghosh, K.~Glaesemann, A.~W. G\"{o}tz, J.~Hammond, V.~Helms,
  E.~D. Hermes, K.~Hirao, S.~Hirata, M.~Jacquelin, L.~Jensen, B.~G. Johnson,
  H.~J'{o}nsson, R.~A. Kendall, M.~Klemm, R.~Kobayashi, V.~Konkov,
  S.~Krishnamoorthy, M.~Krishnan, Z.~Lin, R.~D. Lins, R.~J. Littlefield, A.~J.
  Logsdail, K.~Lopata, W.~Ma, A.~V. Marenich, J.~{Martin del Campo},
  D.~{Mejia-Rodriguez}, J.~E. Moore, J.~M. Mullin, T.~Nakajima, D.~R.
  Nascimento, J.~A. Nichols, P.~J. Nichols, J.~Nieplocha,
  A.~{Otero-de-la-Roza}, B.~Palmer, A.~Panyala, T.~Pirojsirikul, B.~Peng,
  R.~Peverati, J.~Pittner, L.~Pollack, R.~M. Richard, P.~Sadayappan, G.~C.
  Schatz, W.~A. Shelton, D.~W. Silverstein, D.~M.~A. Smith, T.~A. Soares,
  D.~Song, M.~Swart, H.~L. Taylor, G.~S. Thomas, V.~Tipparaju, D.~G. Truhlar,
  K.~Tsemekhman, T.~{Van Voorhis}, A.~V\'{a}zquez-Mayagoitia, P.~Verma,
  O.~Villa, A.~Vishnu, K.~D. Vogiatzis, D.~Wang, J.~H. Weare, M.~J. Williamson,
  T.~L. Windus, K.~Woli\'{n}ski, A.~T. Wong, Q.~Wu, C.~Yang, Q.~Yu,
  M.~Zacharias, Z.~Zhang, Y.~Zhao, R.~J. Harrison, {NWChem}: Past, present, and
  future, The Journal of Chemical Physics 152 (2020) 184102.

\bibitem{octopus_tancognedejean_2020}
N.~Tancogne-Dejean, M.~J.~T. Oliveira, X.~Andrade, H.~Appel, C.~H. Borca,
  G.~{Le Breton}, F.~Buchholz, A.~Castro, S.~Corni, A.~A. Correa, U.~D.
  Giovannini, A.~Delgado, F.~G. Eich, J.~Flick, G.~Gil, A.~Gomez, N.~Helbig,
  H.~H\"{u}bener, R.~Jest\"{a}dt, J.~Jornet-Somoza, A.~H. Larsen, I.~V.
  Lebedeva, M.~L\"{u}ders, M.~A.~L. Marques, S.~T. Ohlmann, S.~Pipolo,
  M.~Rampp, C.~A. Rozzi, D.~A. Strubbe, S.~A. Sato, C.~Sch\"{a}fer,
  I.~Theophilou, A.~Welden, A.~Rubio, Octopus, a computational framework for
  exploring light-driven phenomena and quantum dynamics in extended and finite
  systems, The Journal of Chemical Physics 152 (2020) 124119.

\bibitem{openmx_ozaki_2003}
T.~Ozaki, Variationally optimized atomic orbitals for large-scale electronic
  structures, Physical Review B 67 (2003) 155108.

\bibitem{quantumatk_smidstrup_2019}
S.~Smidstrup, T.~Markussen, P.~Vancraeyveld, J.~Wellendorff, J.~Schneider,
  T.~Gunst, B.~Verstichel, D.~Stradi, P.~A. Khomyakov, U.~G. Vej-Hansen, M.-E.
  Lee, S.~T. Chill, F.~Rasmussen, G.~Penazzi, F.~Corsetti, A.~Ojanper\"{a},
  K.~Jensen, M.~L.~N. Palsgaard, U.~Martinez, A.~Blom, M.~Brandbyge,
  K.~Stokbro, {QuantumATK}: An integrated platform of electronic and
  atomic-scale modelling tools, Journal of Physics: Condensed Matter 32 (2019)
  015901.

\bibitem{qe_giannozzi_2020}
P.~Giannozzi, O.~Baseggio, P.~Bonf\`{a}, D.~Brunato, R.~Car, I.~Carnimeo,
  C.~Cavazzoni, S.~de~Gironcoli, P.~Delugas, F.~F. Ruffino, A.~Ferretti,
  N.~Marzari, I.~Timrov, A.~Urru, S.~Baroni, {Quantum ESPRESSO} toward the
  exascale, The Journal of Chemical Physics 152 (2020) 154105.

\bibitem{vasp_kresse_1996}
G.~Kresse, J.~Furthm\"{u}ller, Efficient iterative schemes for ab initio
  total-energy calculations using a plane-wave basis set, Physical Review B 54
  (1996) 11169--11186.

\bibitem{wien2k_blaha_2020}
P.~Blaha, K.~Schwarz, F.~Tran, R.~Laskowski, G.~K.~H. Madsen, L.~D. Marks,
  {WIEN2k}: An {APW+lo} program for calculating the properties of solids, The
  Journal of Chemical Physics 152 (2020) 074101.

\bibitem{elpa_kus_2019a}
P.~K\r{u}s, H.~Lederer, A.~Marek, {GPU} optimization of large-scale eigenvalue
  solver, in: Numerical Mathematics and Advanced Applications ENUMATH 2017,
  Springer International Publishing, 2019, pp. 123--131.

\bibitem{slate}
\url{https://icl.utk.edu/slate} (Accessed: 2020-09-07).

\bibitem{elpa_kus_2019b}
P.~K\r{u}s, A.~Marek, S.~S. K\"{o}cher, H.-H. Kowalski, C.~Carbogno,
  C.~Scheurer, K.~Reuter, M.~Scheffler, H.~Lederer, Optimizations of the
  eigensolvers in the {ELPA} library, Parallel Computing 85 (2019) 167--177.

\bibitem{elpa_auckenthaler_2013}
T.~Auckenthaler, Highly scalable eigensolvers for petaflop applications, Ph.D.
  thesis, Technische Universit\"{a}t M\"{u}nchen (2013).

\bibitem{lapack_anderson_1999}
E.~Anderson, Z.~Bai, C.~Bischof, L.~S. Blackford, J.~Demmel, J.~Dongarra,
  J.~{Du Croz}, A.~Greenbaum, S.~Hammarling, A.~McKenney, D.~Sorensen, {LAPACK}
  users' guide, SIAM, 1999.

\bibitem{scalapack_blackford_1997}
L.~S. Blackford, J.~Choi, A.~Cleary, E.~D'Azevedo, J.~Demmel, I.~Dhillon,
  J.~Dongarra, S.~Hammarling, G.~Henry, A.~Petitet, et~al., {ScaLAPACK} users'
  guide, SIAM, 1997.

\bibitem{elemental_poulson_2013}
J.~Poulson, B.~Marker, R.~A. {van de Geijn}, J.~R. Hammond, N.~A. Romero,
  Elemental: A new framework for distributed memory dense matrix computations,
  ACM Transactions on Mathematical Software 39 (2013) 13:1--13:24.

\bibitem{mkl_arturov_2018}
K.~Arturov, Intel math kernel library ({Intel MKL}) 2018 update 2: {ScaLAPACK}
  symmetric eigensolver enhancements,
  \url{https://software.intel.com/content/www/us/en/develop/tools/math-kernel-library}
  (Accessed: 2020-09-07).

\bibitem{dc_cuppen_1980}
J.~J.~M. Cuppen, A divide and conquer method for the symmetric tridiagonal
  eigenproblem, Numerische Mathematik 36 (1980) 177--195.

\bibitem{dc_gu_1995}
M.~Gu, S.~C. Eisenstat, A divide-and-conquer algorithm for the symmetric
  tridiagonal eigenproblem, SIAM Journal on Matrix Analysis and Applications 16
  (1995) 172--191.

\bibitem{dc_tisseur_1999}
F.~Tisseur, J.~Dongarra, A parallel divide and conquer algorithm for the
  symmetric eigenvalue problem on distributed memory architectures, SIAM
  Journal on Scientific Computing 20 (1999) 2223--2236.

\bibitem{elpa_gutheil_2014}
I.~Gutheil, J.~F. M\"{u}nchhalfen, J.~Grotendorst, Performance of dense
  eigensolvers on {BlueGene/Q}, in: Parallel Processing and Applied
  Mathematics, Springer Berlin Heidelberg, 2014, pp. 26--35.

\bibitem{elpa_cook_2018}
B.~Cook, T.~Kurth, J.~Deslippe, P.~Carrier, N.~Hill, N.~Wichmann, Eigensolver
  performance comparison on {Cray XC} systems, Concurrency and Computation:
  Practice and Experience (2018) e4997.

\bibitem{householder_householder_1958}
A.~S. Householder, Unitary triangularization of a nonsymmetric matrix, Journal
  of the ACM 5 (1958) 339--342.

\bibitem{mps}
{NVIDIA} multi-process service, \url{https://docs.nvidia.com/deploy/mps}
  (Accessed: 2020-11-09).

\bibitem{cuda_nickolls_2008}
J.~Nickolls, I.~Buck, M.~Garland, Scalable parallel programming, in: 2008 IEEE
  Hot Chips 20 Symposium (HCS), IEEE, 2008, pp. 40--53.

\bibitem{nvlink_foley_2017}
D.~Foley, J.~Danskin, Ultra-performance {Pascal GPU} and {NVLink} interconnect,
  IEEE Micro 37 (2017) 7--17.

\bibitem{cbts_shin_2016}
D.~Shin, B.~Saparov, T.~Zhu, W.~P. Huhn, V.~Blum, D.~B. Mitzi,
  {BaCu2Sn(S,Se)4}: Earth-abundant chalcogenides for thin-film photovoltaics,
  Chemistry of Materials 28 (2016) 4771--4780.

\bibitem{cbts_shin_2017}
D.~Shin, T.~Zhu, X.~Huang, O.~Gunawan, V.~Blum, D.~B. Mitzi, Earth-abundant
  chalcogenide photovoltaic devices with over 5\% efficiency based on a
  {Cu2BaSn(S,Se)4} absorber, Advanced Materials 29 (2017) 1606945.

\bibitem{sic_nemec_2013}
L.~Nemec, V.~Blum, P.~Rinke, M.~Scheffler, Thermodynamic equilibrium conditions
  of graphene films on {SiC}, Physical Review Letters 111 (2013) 065502.

\bibitem{sic_tu_2016}
Q.~Tu, B.~Lange, Z.~Parlak, J.~M.~J. Lopes, V.~Blum, S.~Zauscher, Quantitative
  subsurface atomic structure fingerprint for {2D} materials and
  heterostructures by first-principles-calibrated contact-resonance atomic
  force microscopy, ACS Nano 10 (2016) 6491--6500.

\bibitem{gpu_huhn_2020}
W.~P. Huhn, B.~Lange, V.~W.-z. Yu, M.~Yoon, V.~Blum, {GPGPU} acceleration of
  all-electron electronic structure theory using localized numeric
  atom-centered basis functions, Computer Physics Communications 254 (2020)
  107314.

\bibitem{raw_data}
\url{https://doi.org/10.6084/m9.figshare.13551365.v1} (Accessed: 2021-01-10).

\end{thebibliography}

\end{document}